\documentclass[fleqn,usenatbib]{mnras}

\usepackage{newtxtext,newtxmath}

\usepackage[T1]{fontenc}

\DeclareRobustCommand{\VAN}[3]{#2}
\let\VANthebibliography\thebibliography
\def\thebibliography{\DeclareRobustCommand{\VAN}[3]{##3}\VANthebibliography}

\usepackage{graphicx}	
\usepackage{amsmath}	

\usepackage{amssymb}	

\title[]{Turbulence and its connection to episodic accretion in binary YSOs}

\author[R. Riaz,  D.R.G. Schleicher, S. Vanaverbeke and Ralf S. Klessen]{R. Riaz$^{1}$\thanks{E-mail: rriaz@astro-udec.cl} 
	D.R.G. Schleicher$^{1}$\thanks{E-mail: dschleicher@astro-udec.cl} S. Vanaverbeke$^{2}$\thanks{E-mail: siegfriedvanaverbeke@gmail.com} Ralf S. Klessen$^{3,4}$\thanks{E-mail: klessen@uni-heidelberg.de} \\
	$^{1}$Departamento de Astronom\'ia, Facultad Ciencias F\'isicas y Matem\'aticas, Universidad de Concepci\'on, Av. Esteban Iturra s/n Barrio \\
	Universitario, Casilla $160$-C, Concepci\'on, Chile \\
	$^{2}$Centre for mathematical Plasma-Astrophysics, Department of Mathematics, KU Leuven, Celestijnenlaan 200B, 3001 Heverlee, Belgium \\ 
	$^{3}$Universit{\"a}t Heidelberg, Zentrum f{\"u}r Astronomie, Institut  f{\"u}r Theoretische Astrophysik, Albert-Ueberle-Str. 2,  69120 Heidelberg, Germany \\
	$^{4}$Universit{\"a}t Heidelberg, Interdisziplin{\"a}res Zentrum f{\"u}r Wissenschaftliches Rechnen, Im Neuenheimer Feld 205,  69120 Heidelberg, Germany
}

\date{Accepted 2021 August 26. Received 2021 August 01; in original form 2021 June 22}

\pubyear{2015}

\begin{document}
\label{firstpage}
\pagerange{\pageref{firstpage}--\pageref{lastpage}}
\maketitle


\begin{abstract}
We report signatures of episodic accretion in young stellar objects (YSOs) that emerge in protobinary configurations in a gravoturbulent gas collapse. We find in most of these protobinary systems strong accretion bursts between the two companions with a recurrence time-scale of about 1 kyr. The accretion rate onto the secondary star typically exceeds that onto the primary with a peak value of 2 $\times 10^{-2}$ M$_{\odot}$ yr$^{-1}$ for the former and 6 $\times 10^{-3}$ M$_{\odot}$ yr$^{-1}$ for the latter. We propose that the secondary companion which remains more active in its episodes of accretion bursts, especially for the gas cores with subsonic velocity dispersion, may provide observational opportunities to find traces of episodic accretion in the surrounding gas of the embedded YSOs that are in a binary configuration. Also, protostars evolving as single objects in the same environment show fewer accretion bursts and all together a more steady mass growth history. The prestellar cores with subsonic velocity dispersion exhibit an order of magnitude more intense accretion bursts than in the case of cores with supersonic velocity dispersions. The latter shows the formation of some of the protobinaries in which the primary acts as a more actively accreting companion. This can support these binaries to become systems of extreme mass ratio. Moreover, the YSOs in binary configurations with small semi-major axis $a$ $\approx$ 50 au and high mass ratio $q$ > 0.7 support phases of intense episodic accretion. The eccentricity, however, seems to play no significant role in the occurrence of accretion bursts.
\end{abstract}

\begin{keywords}
accretion discs -- hydrodynamics -- turbulence -- protostars -- low-mass

\end{keywords}



\section{Introduction}

Spatial scales play a vital role in the paradigm of astrophysical turbulence. The turbulence itself can be due to different processes, including blast waves from supernovae \citep{zhang2019numerical, kortgen2016supernova, ibanez2016gravitational, padoan2016supernova, bykov2000nonthermal}, or could be due to the accretion process, which may operate at different scales from the galactic accretion disc to protostellar accretion discs \citep{turner2014transport, klessen2010accretion, begelman2009angular, pringle1981accretion}. 

In molecular clouds (MCs) the turbulent energy is cascading down from large to small scales via interacting eddies. This  property of a turbulent medium, together with the effect of self-gravity, may trigger fragmentation and local collapse, leading to the formation of molecular cores where star-formation takes place \citep{mac2004control, larson1981turbulence, chandrasekhar1951gravitational}. Nonetheless, the formation of dense self-gravitating cores is possible only if the parent clumps are not just transient structures as discussed by \citet{ballesteros2003v}.  However, in molecular cores, the turbulence inside the gas may change its role and provide support against gravity. In the presence of a magnetic pressure the gravitational collapse can be substantially delayed \citep{sanhueza2017massive, kauffmann2013low, mckee2007theory}. The build-up of a magnetic pressure can be a natural outcome of gravity-driven turbulence \citep{xu2020turbulence}. Thus in a turbulent state, the gas in a molecular core can start providing additional turbulent pressure to counter gravity which subsequently affects the star-formation process \citep{federrath2013star,federrath2012star,kirk2007dynamics,leorat1990influence,bonazzola1987jeans}. Nevertheless, in molecular cores, star-formation occurs at various star-formation rates \citep{lu2019star,lada2010star,evans2009spitzer, vazquez2009high}. It is therefore important to comprehend the nature of turbulence and what regulates the process of star-formation inside the dense molecular cores. 

Young stellar objects (YSOs) reside in dense molecular cores \citep{benedettini2018catalogue,hogerheijde1999envelope,lada1993environments}. Previous studies related to YSOs have revealed that  binary systems are the most common stellar configuration in most of the stellar populations \citep{goodwin2010binaries,eggleton2008catalogue,duquennoy1991multiplicity}. Also, it has been suggested that most stellar binary systems are formed as binaries and therefore such systems are not merely the result of stellar encounters that take place during the dynamical evolution of dense stellar systems \citep{goodman1993binary}. In this paper, we perform simulations of collapsing prestellar gas cores with varying levels of turbulence to study the formation of  young binary systems and the subsequent episodic accretion. In our simulations, we find binaries systematically forming in every model.
 Observations have revealed some interesting features associated with YSOs, such as the abrupt accretion and extended outflows that often are correlated \citep{Bally2016, shu2007magnetization, lada2006stellar, soker2003launching, goodson1999jets, goodson1997time}. The so-called luminosity problem depicting the observed inconsistencies related to the luminosity of young low-mass protostars \citep{kenyon1990iras} has caught attention in the community. One of the solutions proposed is episodic accretion associated with YSOs \citep{baek2020radiative,kuffmeier2018episodic,riaz2018episodic, kim2012evidences,stamatellos2012episodic,mckee2010luminosity,vorobyov2009variable,whitworth2003impulsively}.
 The gravoturbulent fragmentation of the MCs can yield dense molecular cores with velocity dispersions indicating subsonic or supersonic gas flows \citep{klessen2005quiescent}. We therefore consider both subsonic and supersonic flows. We search for the occurrence of episodic accretion in protostars formed in our simulations. We also aim to quantify the possible role of turbulence in defining both the intensity and frequency of the accretion bursts occurring in isolated and binary YSOs with specific focus on multiple systems. We present our code and the description of our models in section 2. In section 3, we manifest our results and discuss them. Section 4 is reserved for the conclusions.
         
\section{Code and model description}
We use the smoothed particle hydrodynamics (SPH) technique in our simulations. Our computer code that utilizes the SPH methodology is known as GRADSPH \footnote{Webpage GRADSPH: http://www.swmath.org/software/1046} developed by \citet{vanaverbeke2009gradsph}.

We run a total of sixteen models which are divided into two main sets M1a$-$M8a and M1b$-$M8b. Each set represents a different initial seed that is used to generate the initial turbulent velocity structure inside the spherical gas core. The gas core itself has a total mass $M$ = 5 M$_{\odot}$ with a radius $R$ = 0.027 pc.  In each of our models the initial gas density is constant at $\rho_{\rm i} = 3.8 \times 10^{-18}$~ g cm$^{-3}$. The initial gas density considered for the cores in our models is in agreement with the IRAS Sky Survey of bright cores (L1544 and L1689B) from the region of Taurus–Auriga \citep{kirk2005initial}. We set four different values (8, 10, 12, and 14 K) as the initial temperatures of the gas cores (see Table 1). These selected values are inspired by the observational evidence which takes into account the gas tracers and the spectral energy distributions (SED). For instance, the best kinetic temperature to explain N$_{2}$H$^{+}$ observations is 7$\pm$1 K (and 8 K for N$_{2}$D$^{+}$) inside 5600 au, where the gas appears thermalized with dust \citep{pagani2007depletion}. For the Barnard 68 core the gas is warmer, $\sim$10 K, when probed by NH$_{3}$ \citep{bergin2006thermal}. From Spitzer data, the SEDs of the 14 prestellar cores provide temperature estimates which mostly lie in the range of 10$-$13 K \citep{kirk2007initial}. Also, the column densities for CH$_{3}$OH in the cases of L1512 and L1517B suggest temperatures of the prestellar core in the span of 6$-$14 K \citep{bacmann2016origin}.         

The free fall time for the gas core in all of our simulation models is 30.627 kyr, which is calculated from
\begin{equation} \label{freefalltime}
	t_{\rm ff}=\sqrt{\frac{3 \pi}{32 G \rho_{\rm i}}}.
\end{equation}

The ratio of the rotational energy to the gravitational potential energy of the core is $\beta$ = 0.0785. This parameter remains fixed in each model and is defined by

\begin{equation} \label{beta}
\beta=\frac{R^{3}\omega^{2}}{3 G M}.
\end{equation}
 
Similarly, the ratio of the thermal and turbulent energies ($U_{\rm th}$, $U_{\rm turb}$) to the gravitational potential energy ($\Omega$) is described with the parameters $\alpha_{\rm th}$ and $\alpha_{\rm turb}$. These parameters remain model dependent (see Table 1) and are defined by 

\begin{equation} \label{alpha_th}
	\alpha_{\rm th}=\frac{5 R k T}{2 G M \mu m_{\rm H}},
\end{equation}

\begin{equation} \label{alpha_turb}
\alpha_{\rm turb}=\frac{U_{\rm turb}}{|\Omega|},
\end{equation}
where $\Omega$ = $-$$\frac{3}{5} \frac{GM^{2}}{R}$, $U_{\rm th}$ = $\frac{1}{2}Mv^{2}$, $U_{\rm turb}$ = $\frac{1}{2}M\mathcal{M}^{2}c^{2}$. The other involved quantities are  the angular rotational frequency $\omega$ of the gas core, $G$ the gravitational constant, $M$ the mass of the gas core, $R$ the radius of the gas core, $k$ the Boltzmann constant, $\mu = 2.33$ the mean molecular weight, and $m_{\rm H}$ the mass of the hydrogen atom. In our models, the code uses $G=M=R=1$ as internal dimensionless units.

\begin{figure}
	
	\centering
	\includegraphics[width=\columnwidth]{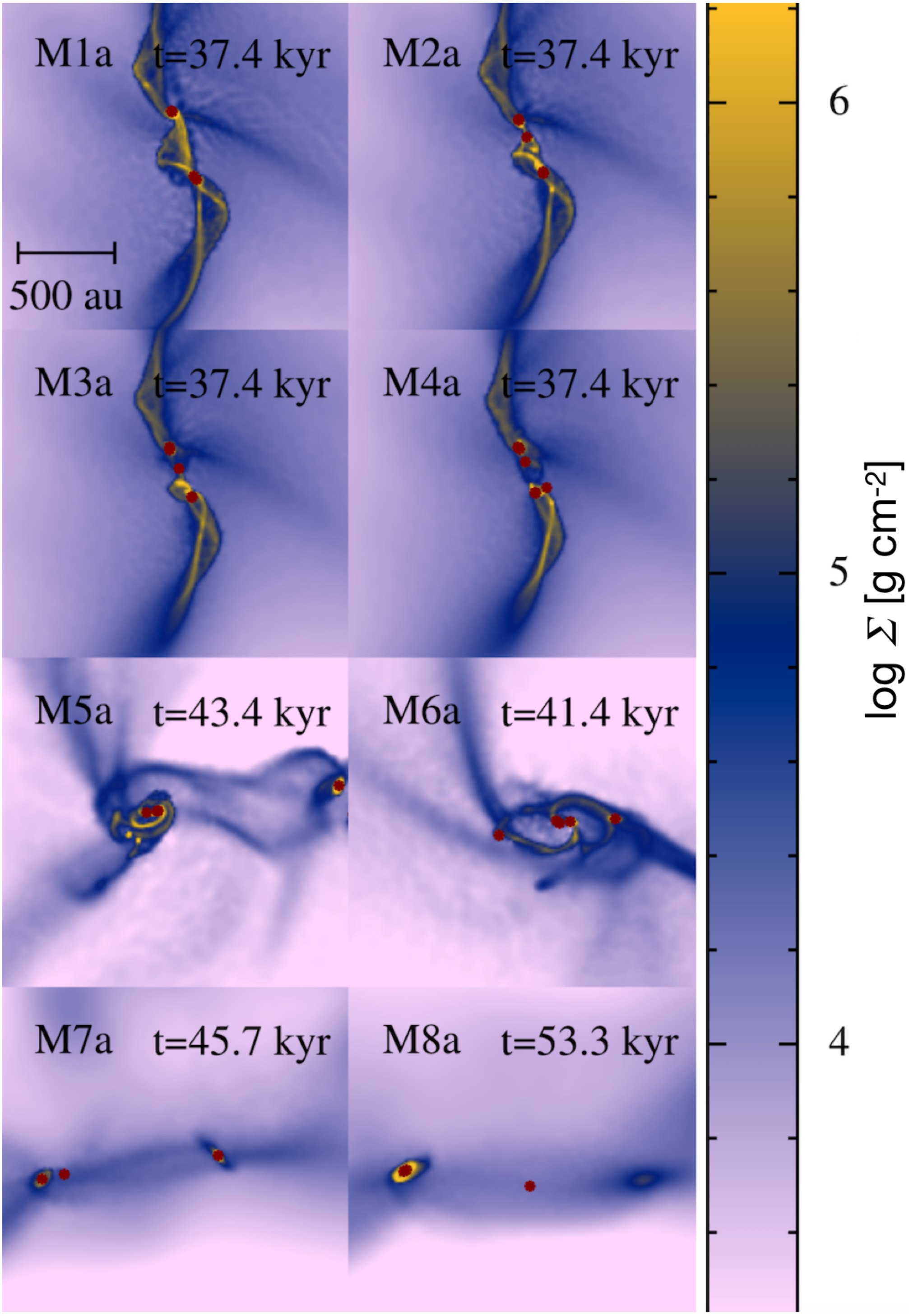}
	\caption{Simulation results for models M1a$-$M8a at the end of the computed evolution of each model. Each panel shows the column density image in the xy-plane. The shaded bar on the right shows column density log ($\rm \Sigma{}$) in g cm$^{-2}$. The corresponding dynamical time in kyr is shown at the top-right corner of each panel. Each calculation was performed with 250025 SPH particles. Colour in online edition.}
	\label{fig:figure1}
\end{figure}

We consider different types of turbulent spectra in our initial conditions, with slopes as appropriate for Kolmogorov \citep{kolmogorov1941dissipation} and Burgers type \citep{burgers1948mathematical} turbulence. The former is incompressible, subsonic turbulence while the latter refers to supersonic, shock-dominated turbulence that can promote formation of dense structures inside the collapsing gas. The scheme implemented to generate the initial velocity structure is described in \citet{riaz2018formation} and was also followed by \citet{riaz2020fragmentation}, hereafter RSVK. We adopt the same scheme in the present work. We inject a spectrum of turbulence into the initial conditions by adding the superposition of the velocity of 1000 shear waves (transverse waves) with random propagating directions to the initial velocity of each particle. The wavelength $\lambda$ of the shear waves is distributed uniformly between $0.001 R$ and $R$, while the amplitude $A$ of the waves follows a spectrum with $A \sim \lambda^{p}$, with the index $p$ taking values of 5/3 and 2.0 in our models for the Kolmogorov and Burgers type turbulence. The amplitude of the resulting turbulent velocity field is then rescaled so that the RMS Mach number of the turbulent flow with respect to the initial isothermal sound speed equals the value in each model. After the generation of the initial conditions, there is no driving of turbulence, but it may dynamically develop, including delay but also driving by infall. In both cases, we consider Mach numbers $\mathcal{M}$ = 0.75 and $\mathcal{M}$ = 3.45, respectively (see Table 1). Our treatment of sink particles (protostars) follows the same method as described in RSVK and the references therein. We set a constant accretion radius $r_{\rm acc}$ = 1 au for the sink particles, which always remains greater than the Jeans length during the gas collapse. A sink particle is introduced inside the collapsing gas whenever the density reaches $10^{-11}$~g cm$^{-3}$. For the merger of sink particles in our simulations, we consider the method as described by \citet{stacy2013constraining}. According to this method, two sinks are allowed to merge if the following three criteria are satisfied:
	\begin{itemize}\setlength{\itemsep}{0.25cm}
		\item When their relative distance $d$ is smaller than $r_{\rm acc}$ (1 au) so that $d < r_{\rm acc}$.
		\item When the total energy $E_{\rm tot}$ of the pair of sink particles is negative, so that the pair is gravitationally bound.
		\item When the least massive sink (secondary) of the pair has insufficient angular momentum to remain rotationally supported against infall onto the massive sink (primary) i.e.  $j_{\rm sec}$ $<$ $j_{\rm cent}$, where $j_{\rm cent}~$=$~\sqrt{G~M_{\rm primary}~d}$ and $M_{\rm primary}$ denotes the mass of the most massive sink of the pair. 
	\end{itemize} 
In our simulations, we use 250025 SPH particles in each model to construct the gas core. For every SPH particle the number of neighboring particles is set as $N_{\rm opt}$ = 50. Thus, following the criterion $M_{\rm resolution}$ = 2 $N_{\rm opt}$ $m_{\rm particle}$, we have in our simulations a minimum resolvable mass $M_{\rm resolvable} = 1.999 \times 10^{-3}$~M$_\odot$.

Our aim is to explore the formation of protostars inside molecular gas cores. These prestellar cores can have a variety of initial thermal and turbulent states \citep{klessen2005quiescent}. We focus on the ability of molecular gas cores to fragment and form protostars, which subsequently accrete material from the surrounding gas. We specifically examine the influence of different spectral slopes corresponding to Kolmogorov and Burgers-type turbulence \citep{bacchini2020evidence,collins2012two,boldyrev2002kolmogorov}. We intend to quantify the effects of these states of turbulence and their effects on the formation and evolution of a protostellar population. The star formation efficiency ($\xi$) is defined as the ratio of the protostellar mass to the gas mass of the parent core. The models M7a, M7b, and M8b required a very high computational cost in our simulations and hence are evolved only up to $\xi$ = 10 \%, both due to the higher Mach number and the higher initial gas temperataure. The rest of the models are followed to $\xi$ = 15 \%. We keep our prime focus on the signatures of episodic accretion in both isolated and binary protostellar configurations that emerge in our simulations. In general, accretion bursts can be the result of a series of frequent protostellar mergers,  effective accretion due to instabilities in circumstellar and circumbinary discs \citep{vorobyov2006burst,vorobyov2005origin}, or even  close encounters in the binary systems \citep{bonnell1992binary}. The publicly available tool for SPH data visualisation SPLASH \citep{price2007splash} is utilized in this work.

Our equation of state to capture the thermodynamical evolution of the collapsing gas core is barotropic of form
\begin{equation} \label{EOS}
P=\rho c_{0}^{2}\left[1+\left(\frac{\rho}{\rho_{\rm crit}}\right)^{\gamma-1}\right],
\end{equation}
where the critical density \textit{$\rho$}$_{\rm crit}$ marks the phase transition from isothermal to adiabatic collapse of the gas. 
\citet{omukai2005thermal} have taken into account the balance between the cooling that is dominated by continuum emission via thermal radiation from the dust and compressional heating. When the gas becomes adiabatic, the relation is  

\begin{equation} \label{transition}
T=\left(\frac{k^{3}}{12\sigma^{2}m_{\rm H}}\right) ^{1/5} {n^{2/5}_{\rm H}},
\end{equation}
where $k$, $\sigma$, $m_{\rm H}$, and $n_{\rm H}$ are the Boltzmann constant, Stefan-Boltzmann constant, mass of the hydrogen atom, and the number density of the gas, respectively. We treat the gas as fully molecular. Since the collapsing gas in our models remains initially in the isothermal phase, the value of \textit{$\rho$}$_{\rm crit}$ which marks the phase transition from isothermal to adiabatic collapse is obtained from equation 6 and is model-dependent (see Table 1). \citet{larson1969numerical} has demonstrated that beyond an approximate density $\sim$ $10^{-13}$~g cm$^{-3}$ the central part of collapsing isothermal gas (10 K) becomes optically thick and the pressure forces increase faster with density compared to the gravitational forces. We use equation (6) to compute the relation between temperature and the critical density \textit{$\rho$}$_{\rm crit}$ at which the EOS of the gas becomes adiabatic (see Table 1).

\begin{table}
	\centering
	\caption{Summary of the initial physical parameters of the simulation models M1a$-$M8a and M1b$-$M8b. The table describes the initial gas temperature ($T_{\rm i}$), the turbulent Mach number ($\mathcal{M}$), the ratio of kinetic energy to the gravitational potential energy of the gas core ($\alpha_{\rm th}$), the ratio of turbulent kinetic energy to the gravitational potential energy of the gas core ($\alpha_{\rm turb}$), and the energy spectral index $p$. For each model, the total mass inside the core, the initial radius of the core, and the initial average number density ($n_{\rm H}$) of the gas core are $5.0$~M$_{\odot}$, $0.027$~ pc, $3.845 \times 10^{-18}$~ g cm$^{-3}$, respectively.} 
	\label{tab:Table1}
	\begin{tabular}{ccccccc} 
		\hline
		\hline
		Model &$T_{\rm i}$ (K)&$\rho_{\rm crit}$~(g cm$^{-3})$& $\mathcal{M}$ & $\alpha_{\rm th}$& $\alpha_{\rm turb}$& index $p$ \\
		\hline
		M1a & 8 &$1.1 \times 10^{-13}$& 0.75 & 0.090& 0.170 & 5/3\\
		M2a & 10 &$1.9 \times 10^{-13}$& 0.75 & 0.113& 0.021& 5/3\\
		M3a & 12 &$3.0 \times 10^{-13}$& 0.75 &0.136 & 0.025& 5/3\\
		M4a & 14 &$4.5 \times 10^{-13}$& 0.75 & 0.159& 0.029& 5/3\\
		M5a & 8 &$1.1 \times 10^{-13}$& 3.45 & 0.090& 0.360& 2\\
		M6a & 10 &$1.9 \times 10^{-13}$& 3.45 & 0.113& 0.450& 2\\
		M7a & 12 &$3.0 \times 10^{-13}$& 3.45 & 0.136& 0.541& 2\\
		M8a & 14 &$4.5 \times 10^{-13}$& 3.45 & 0.159& 0.631& 2\\
		M1b & 8 &$1.1 \times 10^{-13}$& 0.75 & 0.090& 0.170& 5/3\\
		M2b & 10 &$1.9 \times 10^{-13}$& 0.75 & 0.113& 0.021& 5/3\\
		M3b & 12 &$3.0 \times 10^{-13}$& 0.75 & 0.136& 0.025& 5/3\\
		M4b & 14 &$4.5 \times 10^{-13}$& 0.75 & 0.159& 0.029& 5/3\\
		M5b & 8 &$1.1 \times 10^{-13}$& 3.45 & 0.090& 0.360& 2\\
		M6b & 10 &$1.9 \times 10^{-13}$& 3.45 & 0.113& 0.450& 2\\
		M7b & 12 &$3.0 \times 10^{-13}$& 3.45 & 0.136& 0.541& 2\\
		M8b & 14 &$4.5 \times 10^{-13}$& 3.45 & 0.159& 0.631& 2\\
		\hline
	\end{tabular}
\end{table}

\begin{figure}
	
	\centering
	\includegraphics[width=\columnwidth]{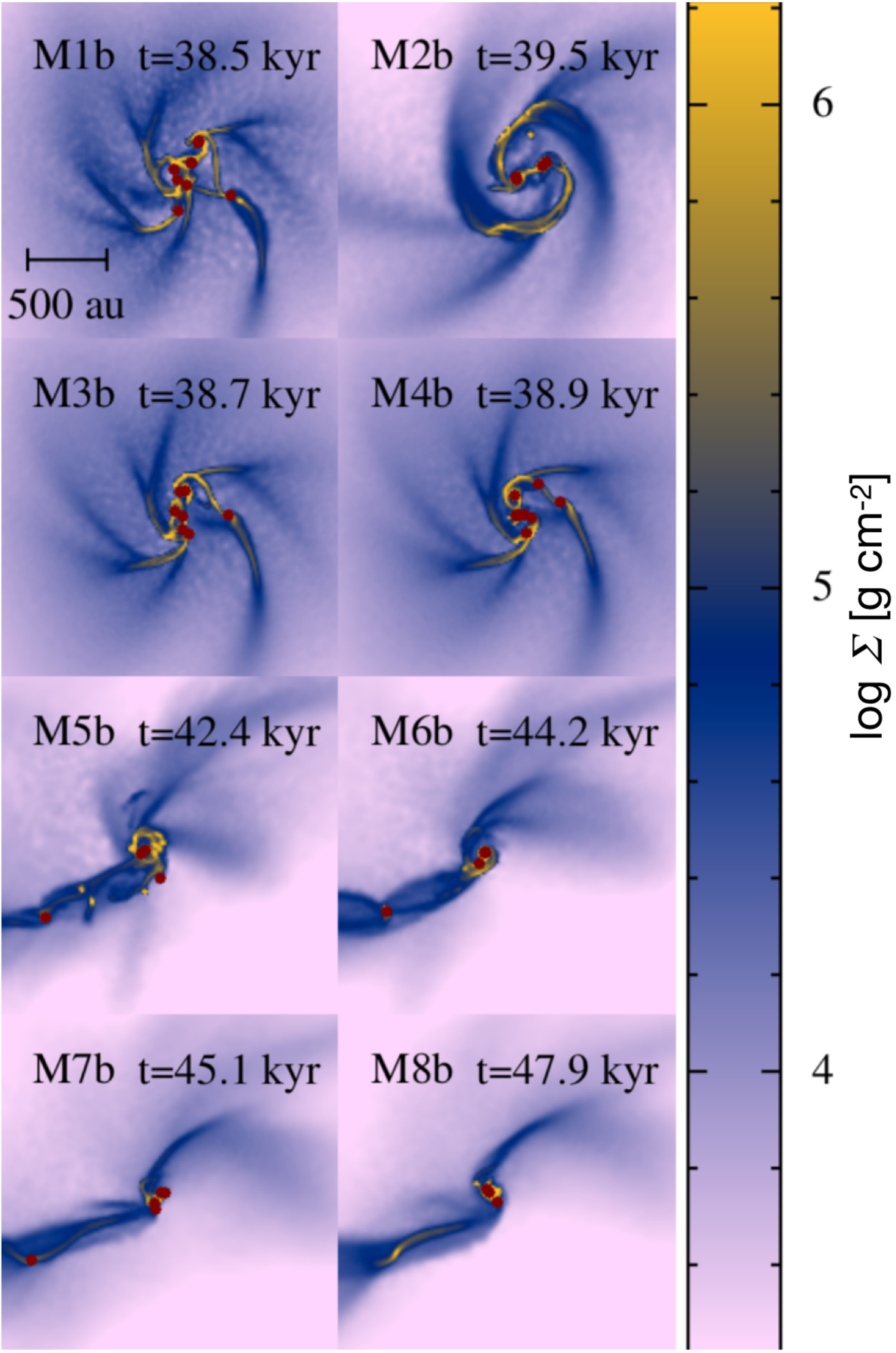}
	\caption{Simulation results for models M1b$-$M8b at the end of the computed evolution of each model. Each panel shows the column density image in the xy-plane. The shaded bar on the right shows column density log ($\rm \Sigma{}$) in g cm$^{-2}$. The corresponding dynamical time in kyr is shown at the top-right corner of each panel. Each calculation was performed with 250025 SPH particles. Colour in online edition.}
	\label{fig:figure2}
\end{figure}

\begin{table}
	\centering
	\caption{Summary of the two sets of models M1a$-$M8a and M1b$-$M8b. The table is constructed at the time when both sets of models reach their final respective states. The table describes the final time ($t_{\rm f}$) where we terminate our simulations, the total number of protostars produced ($N_{\rm max}$) regardless of the merger events, the final number of protostars after the mergers ($N_{\rm proto}$), the binary fraction ($f_{\rm~binary}$), and fraction of the binary contribution towards star formation efficiency ($f_\xi{}_{\rm binary}$).}
	\label{tab:Table2}
	\begin{tabular}{cccccc} 
		\hline
		\hline
		Model & $t_{\rm f}$(kyr) & $N_{\rm max}$ & $N_{\rm proto}$ & $f_{\rm~binary}$ & $f_\xi{}_{\rm binary}$\\
		\hline
		M1a &37.3 &13  & 4 & 1.0 & 1.0\\
		M2a &37.4 &13 & 5 & 0.66 & 0.73\\
		M3a &37.4 &12 & 5 & 0.66 & 0.68\\
		M4a &37.4 &29  & 6 & 0.66 & 0.72\\
		M5a &43.3 &7  & 6 & 0.66 & 0.72\\
		M6a &41.4 &7 & 5 & 0.66 & 0.58\\
		M7a &45.7 &3 & 3 & 0.5 & 0.40\\
		M8a &53.3 &6  & 3 & 0.5 & 0.73\\
		M1b &38.5 &21 & 9 & 0.4 & 0.15\\
		M2b &39.4 &11 & 4 & 0.33 & 0.37\\
		M3b &38.7 &24 & 7 & 0.66 & 0.46\\
		M4b &38.9 &32 & 9 & 0.28 & 0.46\\
		M5b &42.4 &4 & 4 & 0.33 & 0.63\\
		M6b &44.2 &4 & 4 & 0.66 & 0.68\\
		M7b &45.0 &8 & 5 & 0.33 & 0.38\\
		M8b &47.9 &7 & 3 & 0.5 & 0.49\\
		\hline
	\end{tabular}
\end{table}

\begin{table}
	\centering
	\caption{Summary of the final binary properties for the two sets of models M1a$-$M8a and M1b$-$M8b. The columns include the masses of binary components, the binary separation ($d$), the semi-major axis ($a$), the eccentricity ($e$), and the mass ratio ($q$) at the end of the simulation.}
	\label{tab:Table3}
	\begin{tabular}{cccccc} 
		\hline
		\hline
		Model & component masses (M$_{\odot}$) & $d$ (au) & $a$ (au) & $e$ &  $q$    \\
		\hline
		M1a	&	0.21, 0.17	&	34.88	&	38.63	&	0.39	&	 0.80 \\
		M2a	&	0.19, 0.18	&	5.67	&	5.77	&	0.002	&	 0.93 \\
		M3a	&	0.21, 0.15	&	15.11	&	11.34	&	0.40	    &	 0.73 \\
		M4a	&	0.20, 0.14	&	6.32	&	10.43	&	0.39	&	 0.68 \\
		M5a	&	0.28, 0.19	&	7.74	&	7.54	&	0.005	&	 0.68 \\
		M6a	&	0.29, 0.12	&	57.38	&	37.15	&	0.39	&	 0.80 \\
		M7a	&	0.38, 0.002	&	169.3	&	126.8	&	0.57	&	 0.005 \\
		M8a	&	0.42, 0.31	&	16.08	&	17.30	&	0.13	&	 0.73 \\
		M1b	&	0.24, 0.18	&	75.64	&	182.6	&	0.68	&	 0.74 \\
		M2b	&	0.20, 0.17	&	21.95	&	39.20	&	0.55	&	 0.82 \\
		M3b	&	0.20, 0.17	&	67.39	&	74.57	&	0.40	&	 0.86 \\
		M4b	&	0.21, 0.20	&	55.51	&	60.22	&	0.92	&	 0.95 \\
		M5b	&	0.40, 0.22	&	35.72	&	57.45	&	0.38	&	 0.51 \\
		M6b	&	0.28, 0.25	&	9.57	&	8.66	&	0.11	&	 0.89 \\
		M7b	&	0.28, 0.09	&	84.78	&	50.21	&	0.93	&	 0.34 \\
		M8b	&	0.39, 0.09	&	29.65	&	53.96	&	0.45	&	 0.23 \\
		\hline
	\end{tabular}
\end{table}

\section{Results and discussion}
In Figures 1 and 2, we present the morphology of the collapsing gas cores for our simulation models M1a $-$ M8a and M1b $-$ M8b, respectively. The dense filamentary structures appear during the course of gravoturbulent collapse of the gas cores, which remain visible throughout the simulation. They act as the birthplaces of the protostars \citep{dewangan2019evidence,smith2016nature,smith2014nature,hill2011filaments} and often take the shape of spiral arms (especially in the case of second seed). The gravitational collapse results in dense filamentary structures when the turbulent gas is treated to be of Kolmogorov type ($p$ = 5/3). However, for Burgers type ($p$ = 2.0) turbulence, we typically find less concentrated density structures where the spiral/bar shape does not fully develop. The initial Mach number also affects the gas structure during the gravoturbulent collapse. During the gas collapse in dense prestellar cores a rotationally supported structure is formed whose unstable nature triggers the development of strong spiral pattern \citep{goodwin2004simulating,hennebelle2003protostellar,matsumoto2003fragmentation}. For example, \citet{machida2005collapse}, \citet{bate1998collapse} and \citet{durisen1986dynamic} have estimated that the bar-like structure grows only when the ratio of the rotational energy to the gravitational potential energy of the core exceeds $\beta$ = 0.274 . In our models, we find that the incompressible subsonic velocity flows support the formation of spiral and bar-like structures during the gas collapse. However, the models that follow the highly compressible supersonic velocity flows do not favor the formation of well defined spiral/bar-like instabilities. It is important to notice that we use a much weaker rotational factor $\beta$ = 0.0785 in all of our models. 

The morphology of the collapsing gas core affects the subsequent nature of fragmentation in the post-shock gas. In Figures 1 and 2, the enhanced density contrast depends more strongly on the initial turbulent velocity fields of the energy spectra (as defined here by the spectrum index $p$) than on the initial thermal states of the gas cores. This dictates the number of protostars that form during the gas collapse. In general, a greater number of protostars forms inside multiple spiral structures, whereas in case of the more diffused gas structures the number remains smaller (see Table 2).       

\begin{figure}
	
	\centering
	\includegraphics[width=\columnwidth]{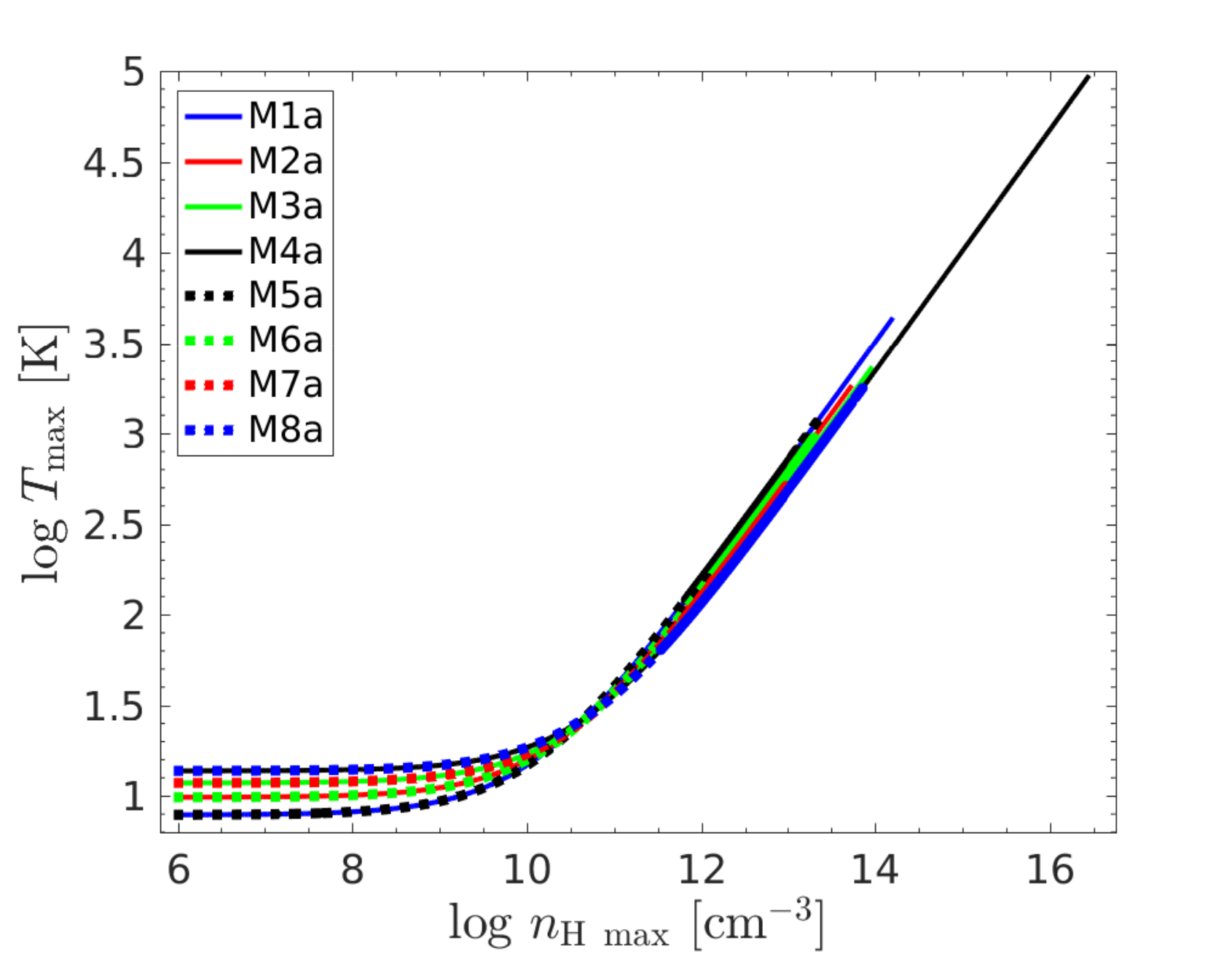}
	\includegraphics[width=\columnwidth]{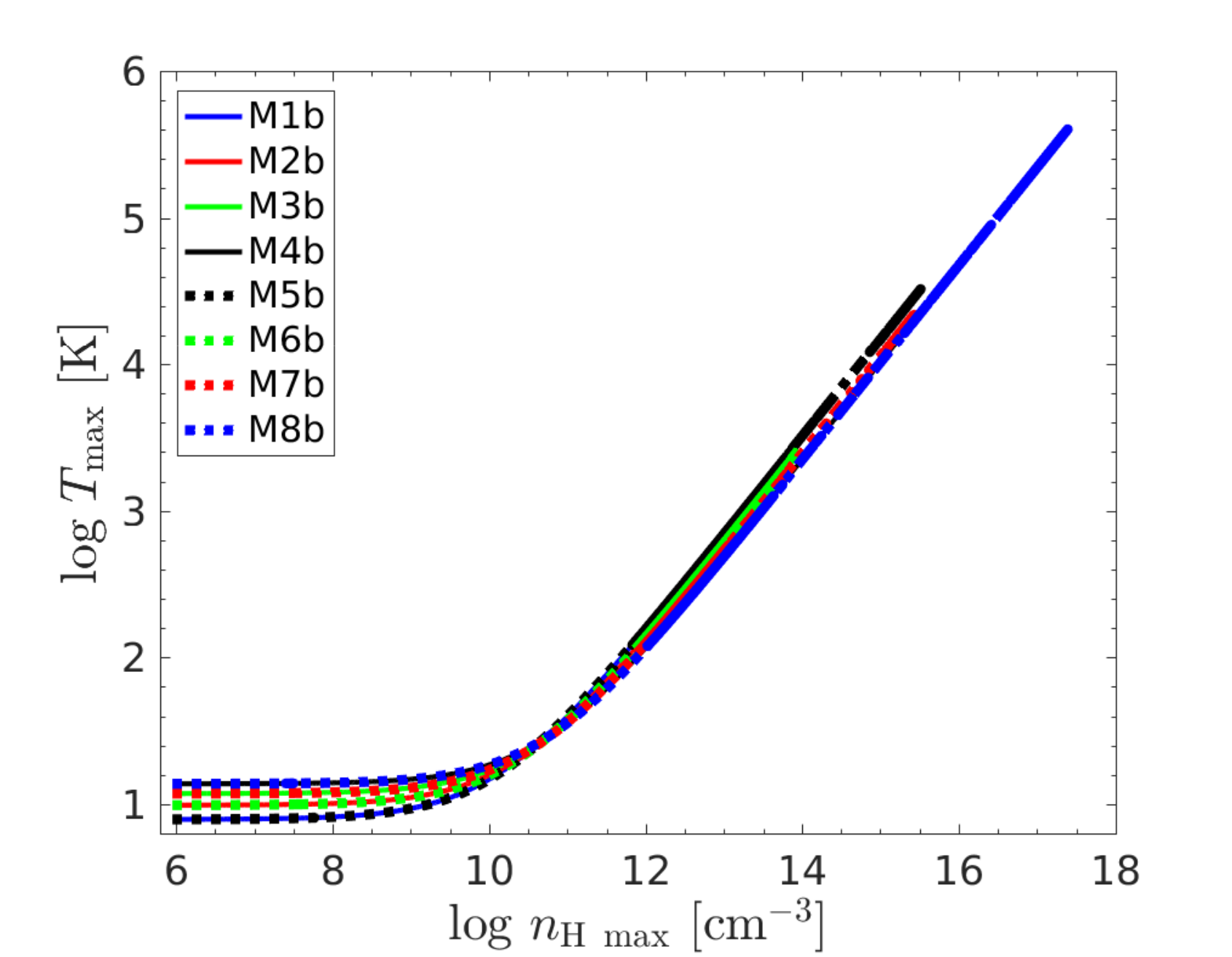}
	\caption{Evolution of the maximum temperature of the gas core as a function of its evolving maximum density during the collapse of the models M1a$-$M8a (top-panel) and models M1b $-$M8b (bottom-panel). The temperature is given Kelvin while the number density is in the unit of cm$^{-3}$. Colour in online edition.}
	\label{fig:figure3}
\end{figure}

\begin{figure}
	
	\centering
	\includegraphics[width=\columnwidth]{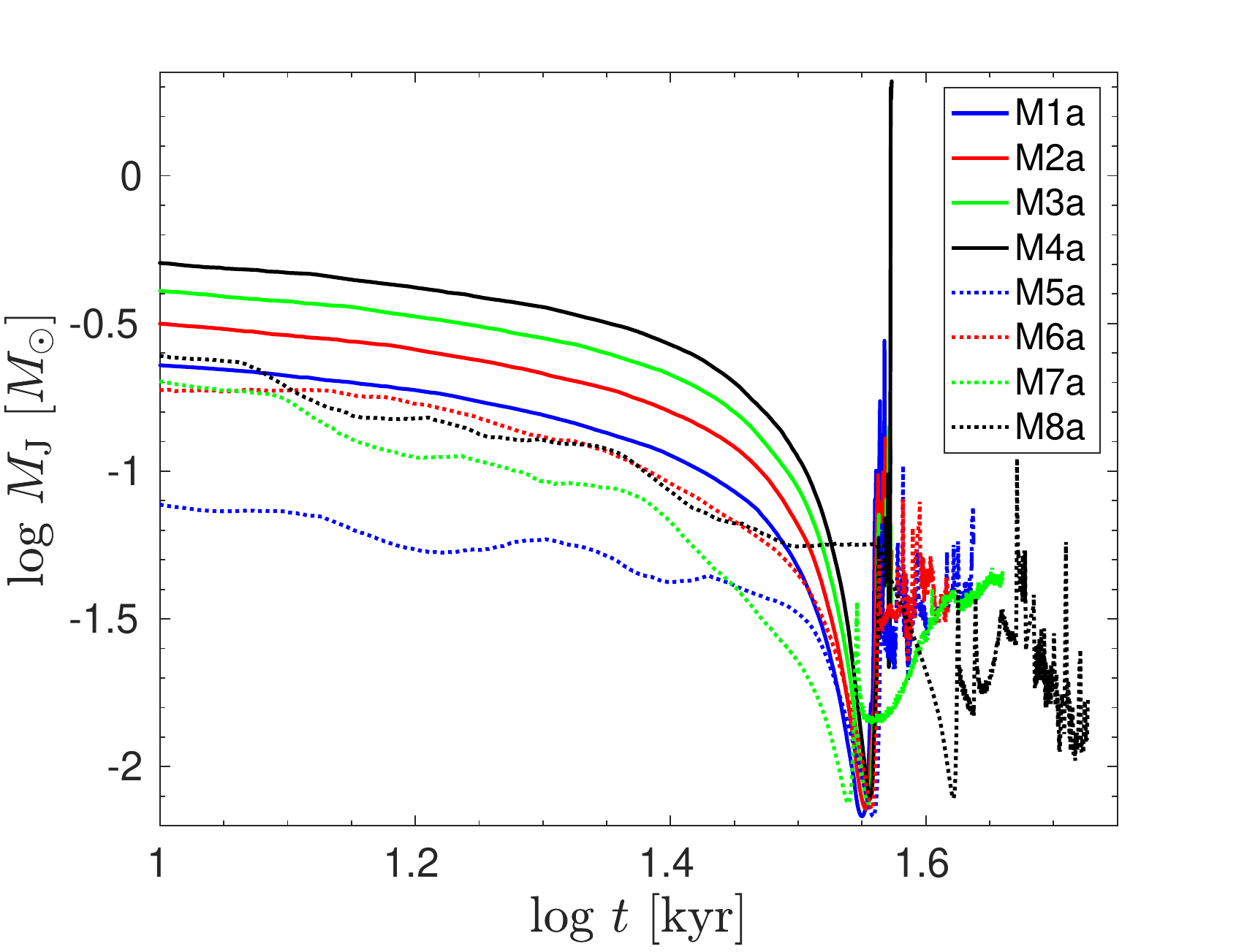}
	\includegraphics[width=\columnwidth]{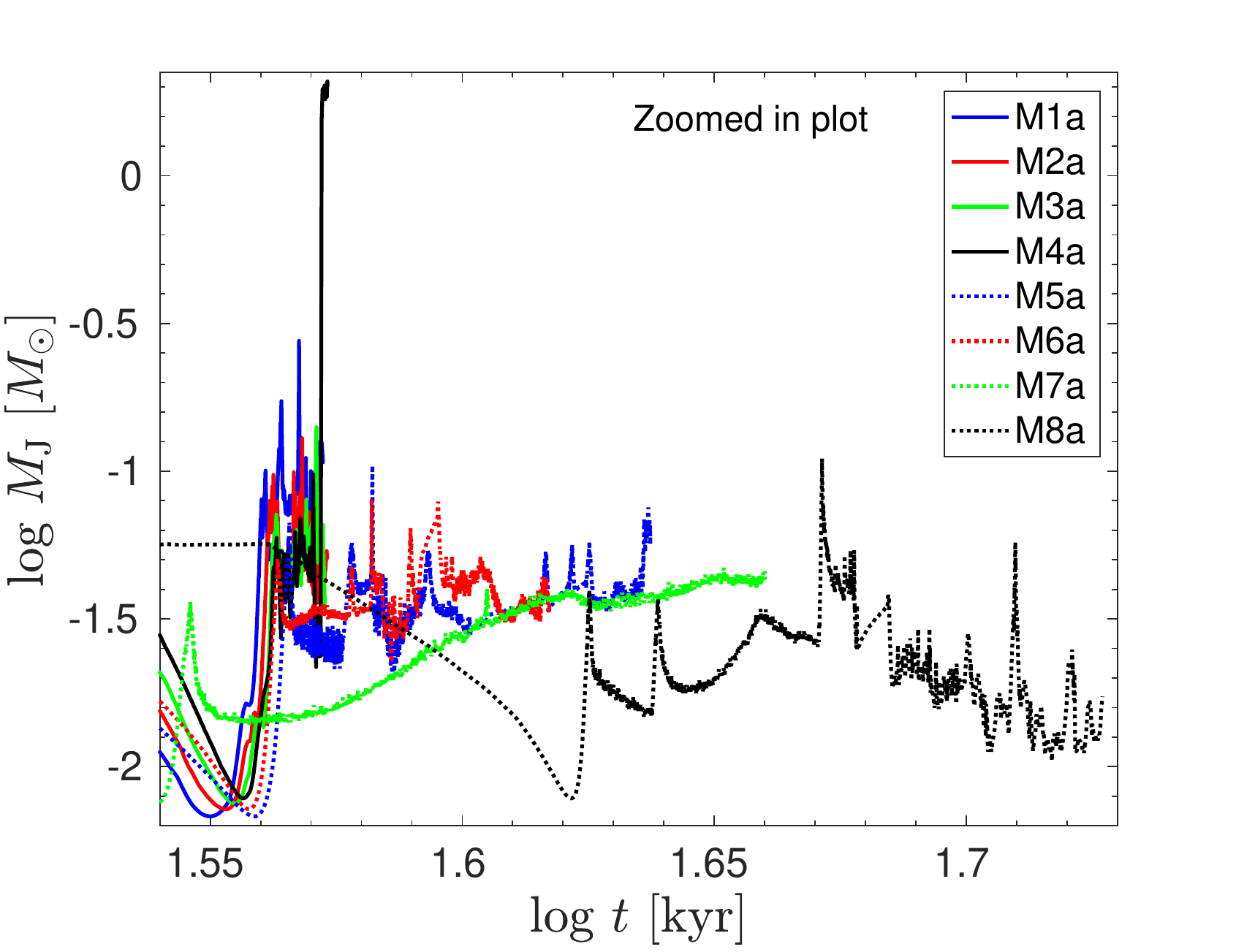}
	\caption{Evolution of the Jeans mass of the gas core as a function of time during the collapse of the models M1a$-$M8a. The top panel provides a view in totality while the bottom panel gives a zoomed-in version. The Jeans mass is given in units of solar mass units while time in kyr. Colour in online edition.}
	\label{fig:figure4}
\end{figure}

\subsection{Thermal response}
Figure 3 shows the overall thermal response of the collapsing gas cores as a function of the increasing gas density. The top and the bottom panels indicate the model results for M1a $-$ M8a and M1b $-$ M8b, respectively. In both cases the gas in the initial phase of collapse remains isothermal as we assume that the radiation generated via gravitational contraction escapes via radiation. However, the increasing gas density due to the collapse becomes opaque to the internal radiation when reaching the critical density the gas, and the core begins to heat up. This behaviour was well studied in  previous attempts when the barotropic EOS was used to model the prestellar core collapse \citep{ bate2016dynamics,whitehouse2006thermodynamics,boss2004fragmentation}.

In the subsonic gas models M1a$-$M4a and M1b$-$M4b, where the initial thermal state is set to vary between 8 K $-$ 14 K, the collapse originally forms a greater number of protostars for the warmer initial thermal states. Contrary to this, the supersonic gas models M5a$-$M8a and M5b$-$M8b (which follow the identical range of initial temperatures) do not exhibit a similar trend. In the latter case, the number, overall, remains small and also dependent on the seed. We believe that modelling the gravoturbulent gas collapse with a larger number of initial seeds can establish a more statistically sound understanding, and provide more clarity about whether the strong initial turbulence plays a vital role and suppresses any potential effect which the initial thermal state of the gas can impart on controlling the process of fragmentation (see Table 2). It may appear counter-intuitive in the context of the thermal Jeans mass condition, however, \citet{padoan1995supersonic} has described that the formation of protostars does not depend only on the mean density and temperature of the environment, but also on the turbulent velocity dispersion and the chemistry responsible for post-shock cooling. Our findings indicating the turbulence as an influential aspect in the context of fragmentation are also consistent with the more recent studies by \citet{booth2019characterizing} and \citet{baehr2017fragmentation}. The effect of turbulence in collapsing gas cores can indeed dominate over the effect from the thermal state of the gas. We find good evidence that in cores with supersonic velocity dispersion there is a strong counterbalance force against gravitational contraction, which subsequently suppresses fragmentation. This also affects the final protostellar masses in our models.  

The Jeans mass $M_{\rm J}$ is given as
\begin{equation} \label{transition}
M_{\rm J}= \left(\frac{5 R_{\rm g} T}{2 \mu G} \right) ^{3/2} \left(\frac{4}{3} \pi \rho \right) ^{-1/2},
\end{equation}
where $R_{\rm g}$ is the ideal gas constant and $T$ is the gas temperature. The gas core heats up and becomes dense during the phase of collapse. Equation 8 is used to analyse the $M_{\rm J}$ as the density and the temperature of the gas evolves with time where the constant in the equation is derived in cgs units, that is $\rho$ and $T$ are measured in g cm$^{-3}$ and K, respectively.
\begin{equation} \label{transition}
M_{\rm J}=2.389779 \times10^{22}~\mathrm{erg~ K^{-1}cm^{-3}s^{2}}~ \rho^{-1/2}~T^{3/2}.   
\end{equation}    

In Figures 4 and 5, we present the thermal Jeans mass $M_{\rm J}$ as a function of time for the models M1a $-$ M8a and M1b $-$ M8b, respectively. The results shown in the top and the bottom panels of these figures indicate  the zoomed-out and zoomed-in versions. The thermal Jeans mass exhibits two distinct phases of isothermal and adiabatic gas collapse \citep{riaz2020fragmentation,omukai2005thermal} when the phase plot shown in Figure 3 is compared with the time evolution of $M_{\rm J}$ in Figures 4 and 5. The isothermal phase of gas collapse lasts until t $\approx$ 0.161 kyr. In the subsequent phase the gas collapses adiabatically. The EOS (here equation 5) indicates the two phases of gas collapse when the adiabatic index $\gamma$ shifts from 1 to 1.6667. During the first phase, $M_{\rm J}$ is a decreasing function of time. In the second phase, $M_{\rm J}$ reverses its behavior due to the ever increasing opacity of the medium as the collapse proceeds in time. We find that the former, irrespective of the initial thermal states of the gas, allows the gas to cool down and attain $M_{\rm J}$ = 0.006 M$_{\odot}$ as the minimum thermal Jeans mass for possible fragmentation. This is an indication that the minimum mass of a possible fragment is not dependent of the initial thermal state of the collapsing gas core. The subsequent phase indicates a rather fluctuating response where the rise in $M_{\rm J}$ shows some dependence on the thermal state of the gas. Nonetheless, it is still consistent with the idea proposed by \citet{klessen2000gravitational} that the supersonic turbulence which remains strong enough to support a molecular cloud against gravity may still allow local collapse to occur (in our case producing fewer number of fragments for supersonic gas collapse when compared with the collapse of subsonic gas). 

\subsection{Mass accumulation}
The total mass accumulated by all the protostars formed during the gravoturbulent gas collapse models is shown in Figure 6. The panel covers model sets M1a $-$ M8a and M1b $-$ M8b in the top and bottom panels, respectively. A generally observed feature in these models is the fraction of gas converted into the protostars (already mentioned as $\xi$). The collapsing gas cores of subsonic velocity flows show a much steeper evolution of the gas-mass converted into the protostars. However, the gas cores of supersonic velocity flows, especially during their later part of evolution, take additional time to reach $\xi = 15 \%$. The subsonic cores offer less turbulent energy support to stabilize the gas against gravitational collapse. This results in more fragmentation in these models. Subsequently, the frequent formation of protostars causes a rapid growth in the star formation efficiency $\xi$ in these models. In contrast, the growth in $\xi$ takes relatively more time in the gas cores with supersonic velocity flows. This is primarily due to the presence of stronger turbulent energy support. 

The individual mass-growth of the protostars via competitive accretion is a common phenomenon in star-forming environments which has consequences on the masses of the protostars \citep{ clark2021emergent,myers2014protostar,veltchev2011stellar,myers2009distribution,clark2008conditions,offner2008driven,bonnell1997accretion}. In subsonic gas models there is more fragmentation occurring, and so there is always a possibility that the  masses of newly formed protostars remain low due to the competitive mass accretion. However, in a densely populated cluster,  the frequent mergers of protostars at the early stage of evolution can still allow protostars to grow their mass, but at the cost of reducing eventually the number of protostars in a cluster.    
We suspect that in addition to the nature of the turbulent velocity flows, the competitive accretion also contributes significantly towards the less massive protostars produced in subsonic gas models. These models though provide potentially more merger events among protostars. Nontheless, our results suggest that at the end of the simulations the final masses of the protostars in the subsonic velocity flows remain smaller. In models with supersonic velocity flows the protostellar mass-growth though remains slower but these models produce more massive protostars as the final product (see Table 3). This can be explained by the inefficient competitive mass accretion process in supersonic gas cores.         

\begin{figure}
	
	\centering
	\includegraphics[width=\columnwidth]{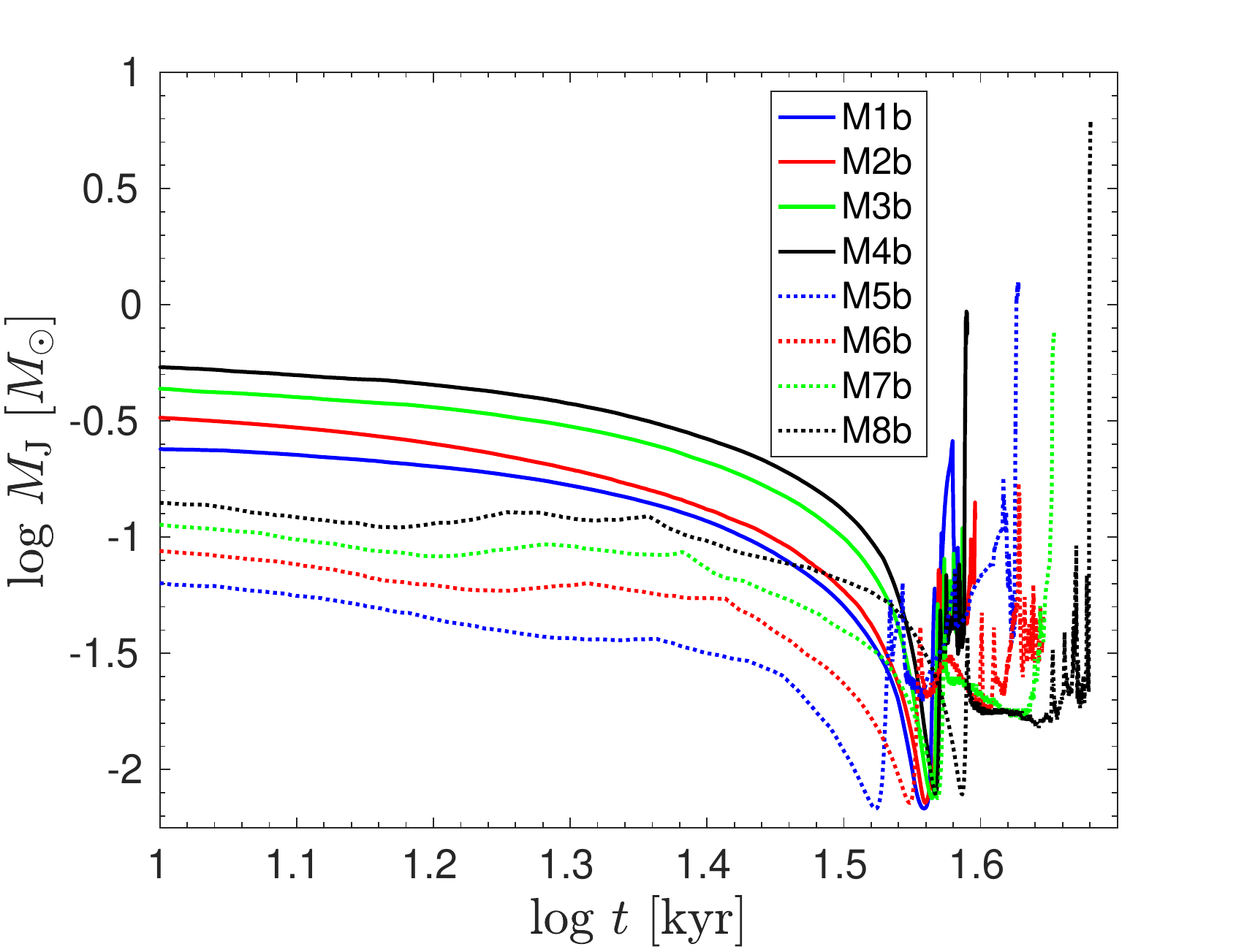}
	\includegraphics[width=\columnwidth]{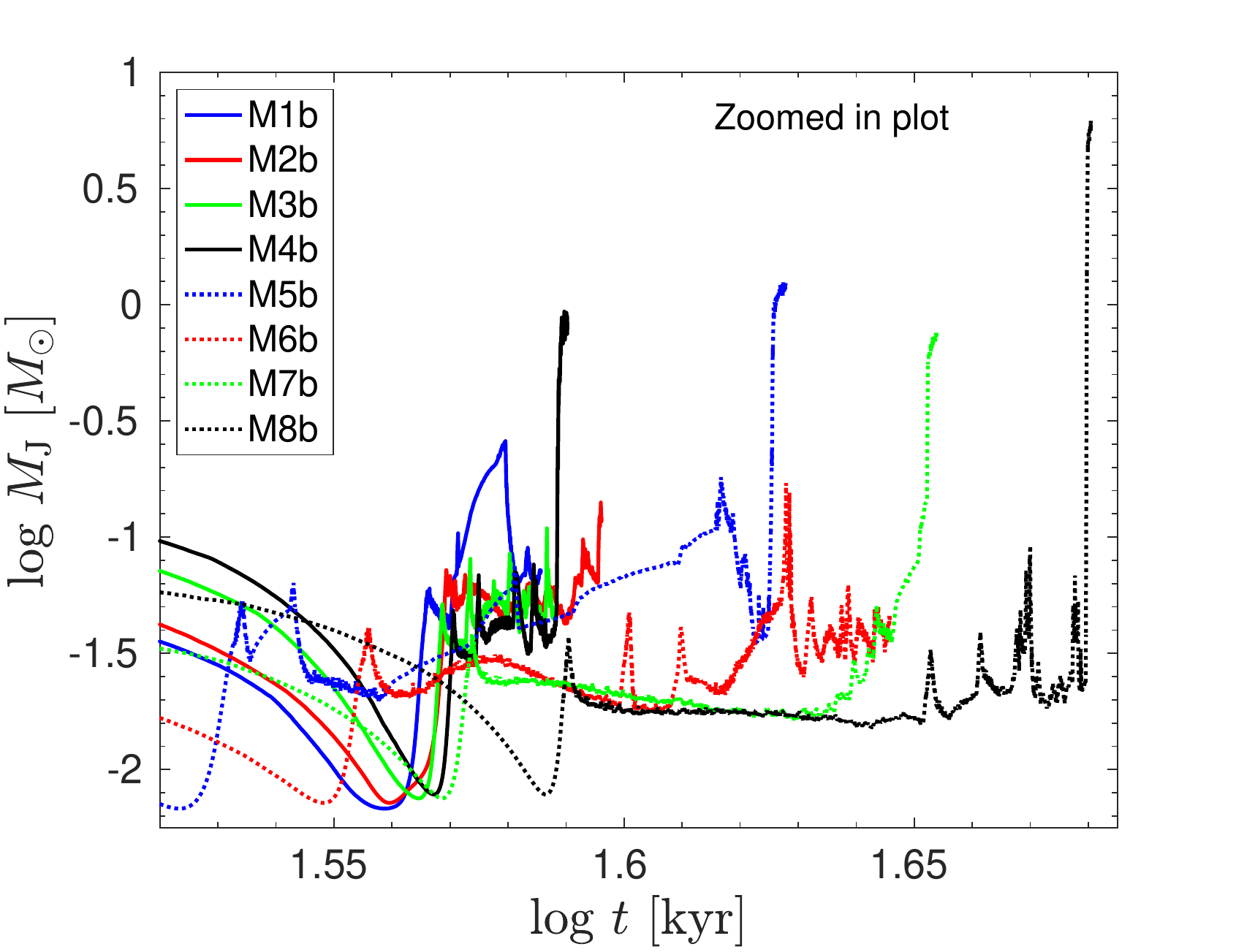}
	\caption{Evolution of the Jeans mass of the gas core as a function of time during the collapse of the models M1b$-$M8b. The top panel provides a view in totality while the bottom panel gives a zoomed-in version. The Jeans mass is given in units of solar mass while time in kyr. Colour in online edition.}
	\label{fig:figure1}
\end{figure}

\begin{figure}
	
	\centering
	\includegraphics[width=\columnwidth]{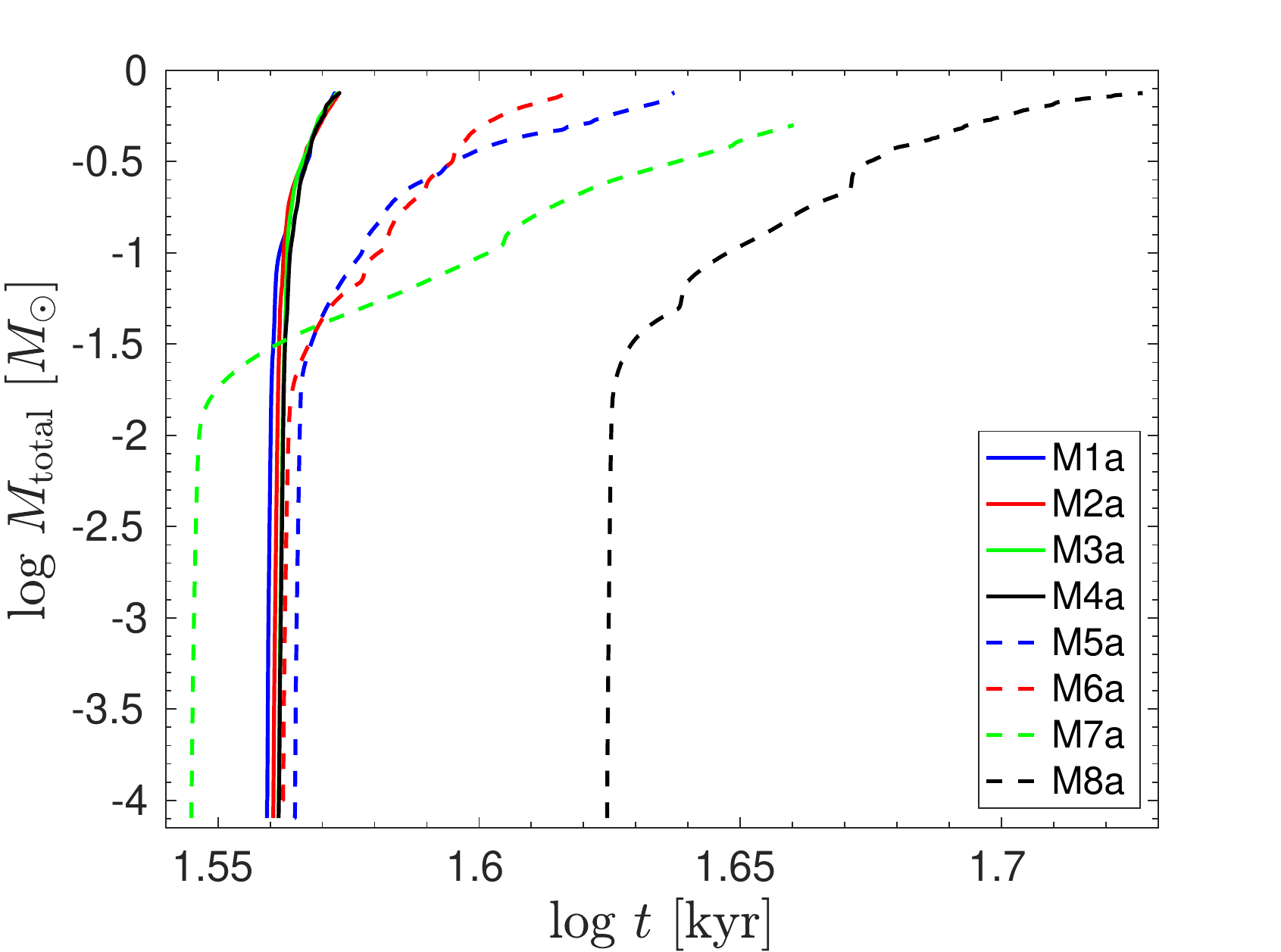}
	\includegraphics[width=\columnwidth]{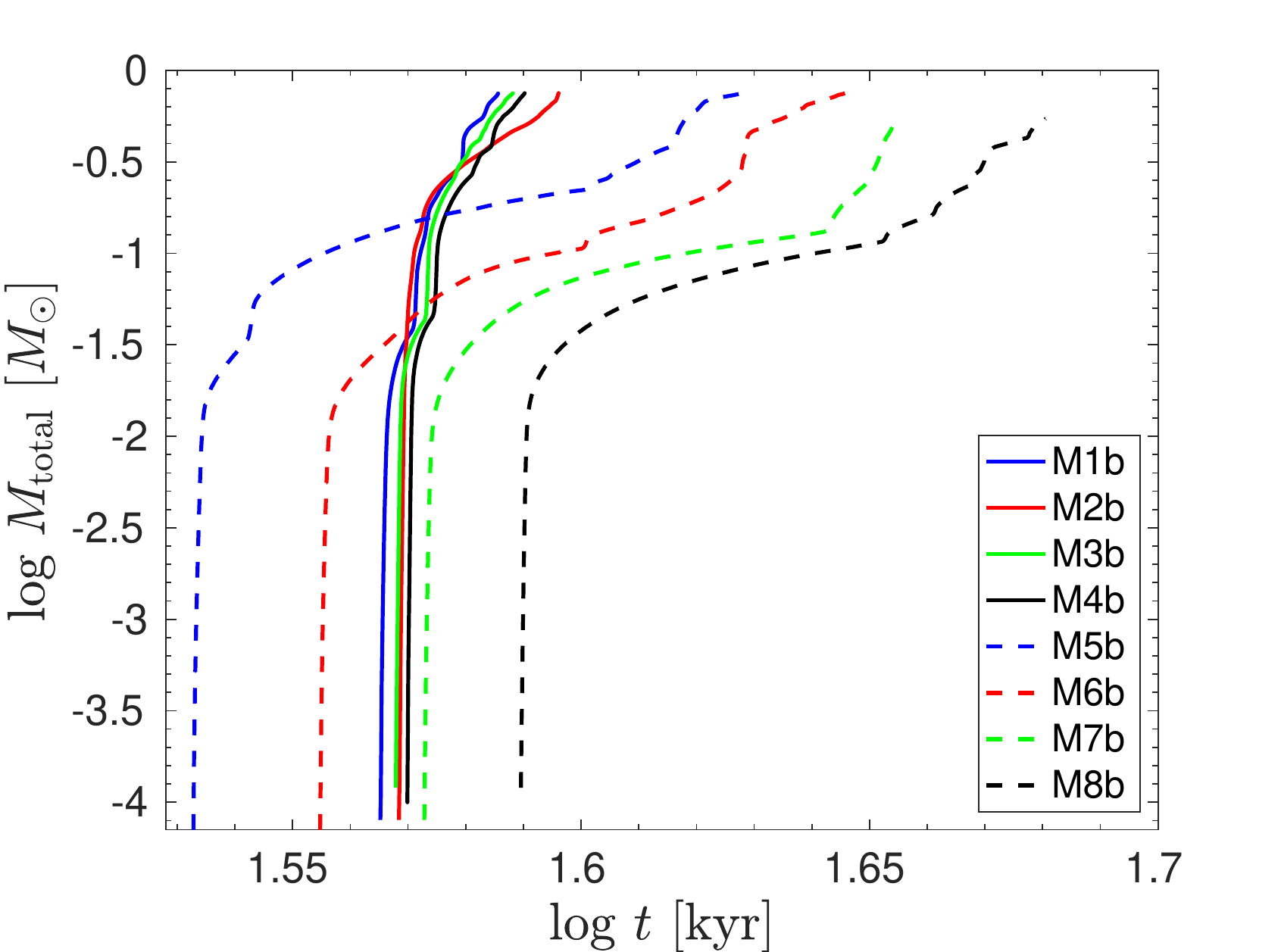}
	\caption{The total protostellar mass accumulation history of all the models M1a$-$M8a (top panel) and M1b$-$M8b (bottom panel) at the end of their simulations. The total gas mass converted into protostars is given in units of solar mass and the time in kyr. Colour in online edition.}
	\label{fig:figure1}
\end{figure} 

\subsection{Accretion rate}
In the previous section, we have shown that simulations with Kolmogorov-type turbulent spectral slopes produce frequent fragmentation. However, the resulting protostars remain small in their masses. We find roughly the opposite behavior in simulations with a Burgers-type slope. Our simulation runs with the second seed confirm a similar emerging trend and make the analysis statistically more sound, though one may need further simulations to draw definite conclusions. 

We now focus on the mass accretion rates of the most massive protostars (isolated and binary components) in the two sets of models M1a $-$ M8a and M1b $-$ M8b and look for  possible signs of episodic accretion expected in YSOs \citep{hsieh2019chronology}. The first spike in mass accretion history illustrated in Figures 7, 8, and A1$-$A6 refers only to the formation event of a protostar when a sink particle is introduced inside the collapsing gas once the gas attains sink-formation density. Therefore, the first spike is not treated as an indicative of an accretion burst \citep{stacy2010first, martel2006fragmentation, bromm2002formation}. The most massive protostars typically formed relatively early during the core collapse. They gain mass both via gas accretion from the surroundings as well as from protostellar mergers. Our choice to select only the most massive protostars thus provides a better opportunity to find and analyse the expected signal of episodic accretion, especially the strong accretion bursts resulting from the efficient gas accretion and merger events. 

For the protobinary systems, our definition of the primary and secondary components is based on their final masses at the time we terminate our simulations. While the less massive companion is considered as a secondary component of the system even if it appears in the gas earlier than the primary. 
The panels in Figures 7 and 8 are arranged to provide a comparison of the mass accretion rates of the most massive protostars evolving under subsonic and supersonic velocity flows for models of a given initial thermal state. For a similar comparison of rest of the models, we provide a detail description of evolution of mass accretion rates in Appendix-A.
It is important to mention here that the very first accretion burst that the two companions of the protobinary system exhibit may not necessarily be strictly related with the binary system. In fact, the first accretion burst can be a feature of the protostar formation itself before they actually form a binary system during the model evolution (as discussed earlier).

Figures 7 shows the evolution of the mass accretion rate of the most massive protobinary system and the most massive isolated protostar (if present) in respective models. In model M1a with Kolmogorov-type turbulence (top panel), the primary component shown in blue forms at $t$ = 36.308 kyr while the secondary in red forms at $t$ = 36.224 kyr. After their formation the two components evolve further in time and beyond $t$ = 36.8 kyr exhibit successive accretion bursts with increasing peak values as high as $\dot M_{\rm prim}$ = 1.0 $\times 10^{-3}$ M$_{\odot}$ yr$^{-1}$ and $\dot M_{\rm sec}$ = 1.7 $\times 10^{-3}$ M$_{\odot}$ yr$^{-1}$, respectively. The successive accretion bursts from both the primary and secondary components continue to occur until we terminate the simulation at $t$ = 37.4 kyr. No isolated protostar is formed in model M1a. 

In the bottom panel of Figure 7, we show the evolving $\dot M$ for model M5a in case of a Burgers-type turbulence. Blue and red indicate the $\dot M$ evolution for the primary and the secondary components of the protobinary system. These two components form at $t$ = 36.643 kyr and at $t$ $\approx$ 37.670 kyr, respectively. After their formation, the two components show the first intense accretion bursts at $t$ = 38.194 kyr and the corresponding peaks in $\dot M$ reach $\dot M_{\rm prim}$ = 9.7 $\times 10^{-4}$ M$_{\odot}$ yr$^{-1}$ and $\dot M_{\rm sec}$ = 8.2 $\times 10^{-4}$ M$_{\odot}$ yr$^{-1}$. From this point in time until $t$ $\approx$ 43.151 kyr the protobinary system keeps showing frequent accretion bursts such that the secondary remains relatively more active than the primary component. This is indicative of high specific angular momentum gas infall into the parestron distance from the centre of mass of the binary system hence making it easier for the secondary component to accrere more than the primary \citep{bate1997effects}. During the later stage of evolution, the protobinary system at $t$ $\approx$ 43.3 kyr shows another intense accretion burst with the peak values of $\dot M_{\rm prim}$ = 1.8 $\times 10^{-4}$ M$_{\odot}$ yr$^{-1}$ and $\dot M_{\rm sec}$ = 6.2 $\times 10^{-4}$ M$_{\odot}$ yr$^{-1}$. The isolated massive protostar shown in green forms at $t$ = 42.072 kyr and does not exhibit any signs of episodic accretion. Its accretion rate continues to drop over time and becomes as small as $\dot M_{\rm iso}$ = 1.0 $\times 10^{-5}$ M$_{\odot}$ yr$^{-1}$ until we terminate the simulation at $t$ = 43.4 kyr. However, the accretion rate $\dot M$ of isolated protostars during the rest of the simulation is decreasing by up to two orders of magnitude. There are no signs of any accretion burst  taking place at the isolated protostar.

\begin{figure}
	\centering
	\includegraphics[width=\columnwidth]{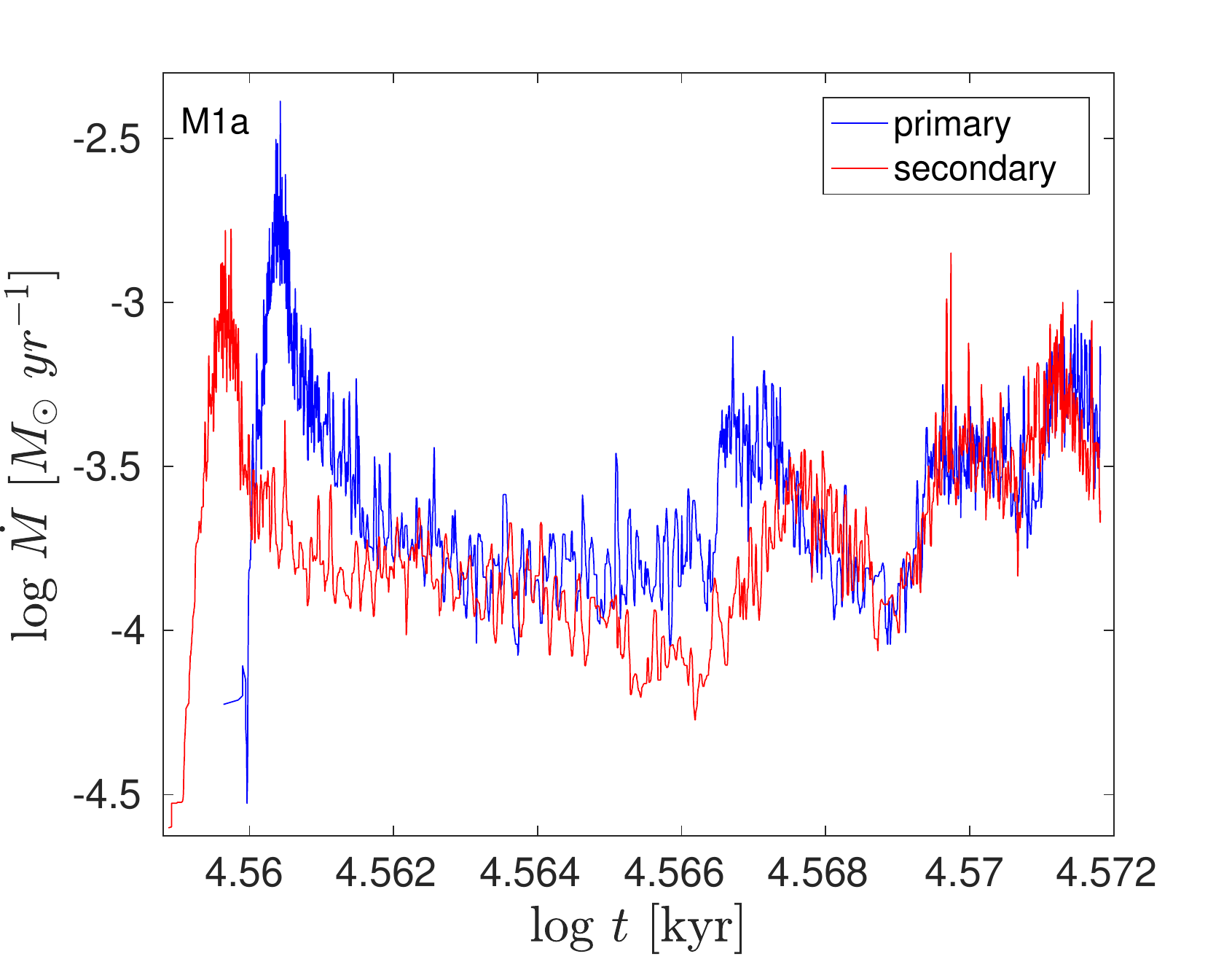}
	\includegraphics[width=\columnwidth]{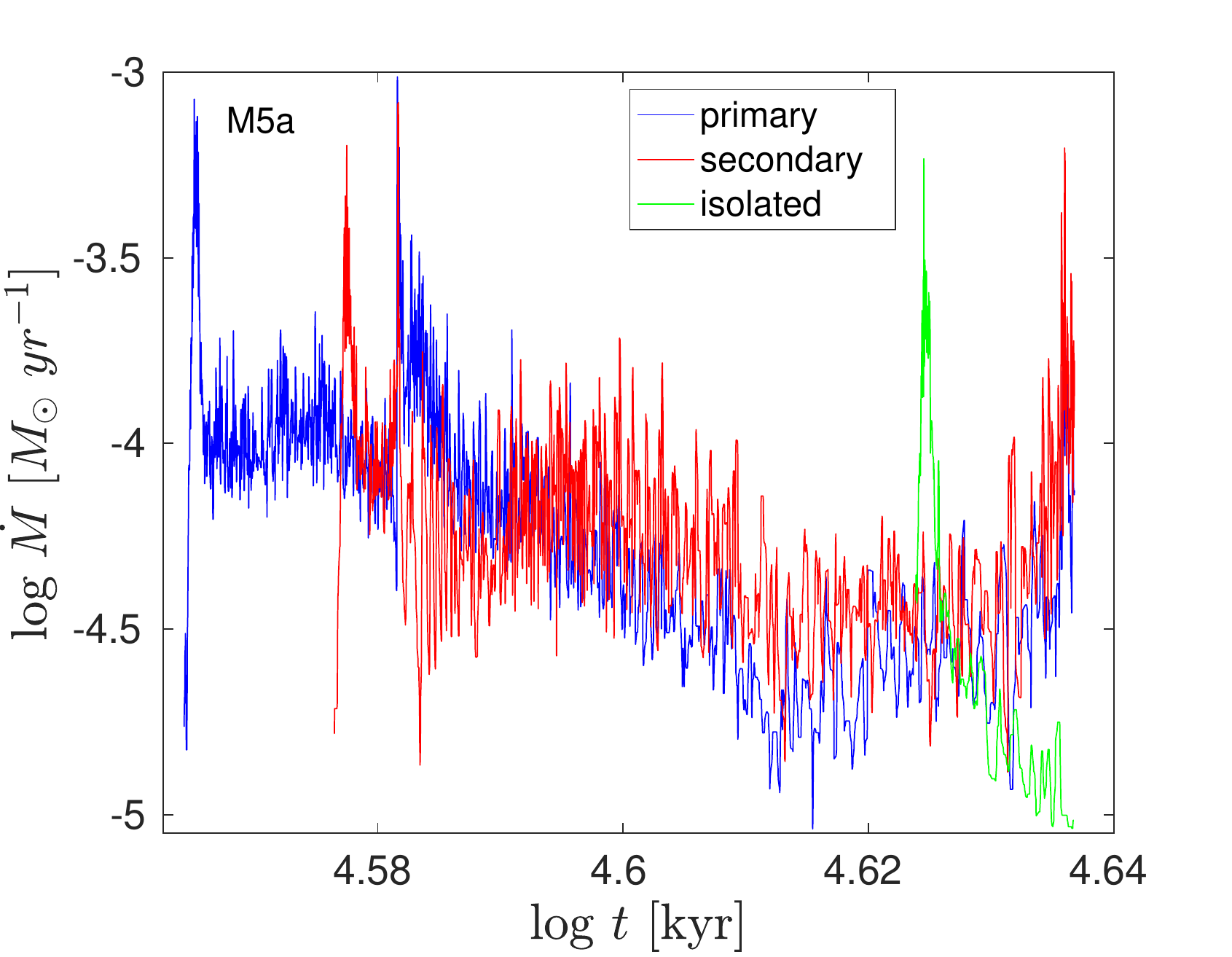}
	\caption{Accretion rates for the primary (blue), secondary (red), and isolated protostars (green) (if present) in models M1a (top panel) and M5a (bottom panel). The accretion rate is given in units of M$_{\odot}$ yr$^{-1}$ and the time is in kyr. Colour in online edition. }
	\label{fig:figure1}
\end{figure}

\begin{figure}
	
	\centering
	\includegraphics[width=\columnwidth]{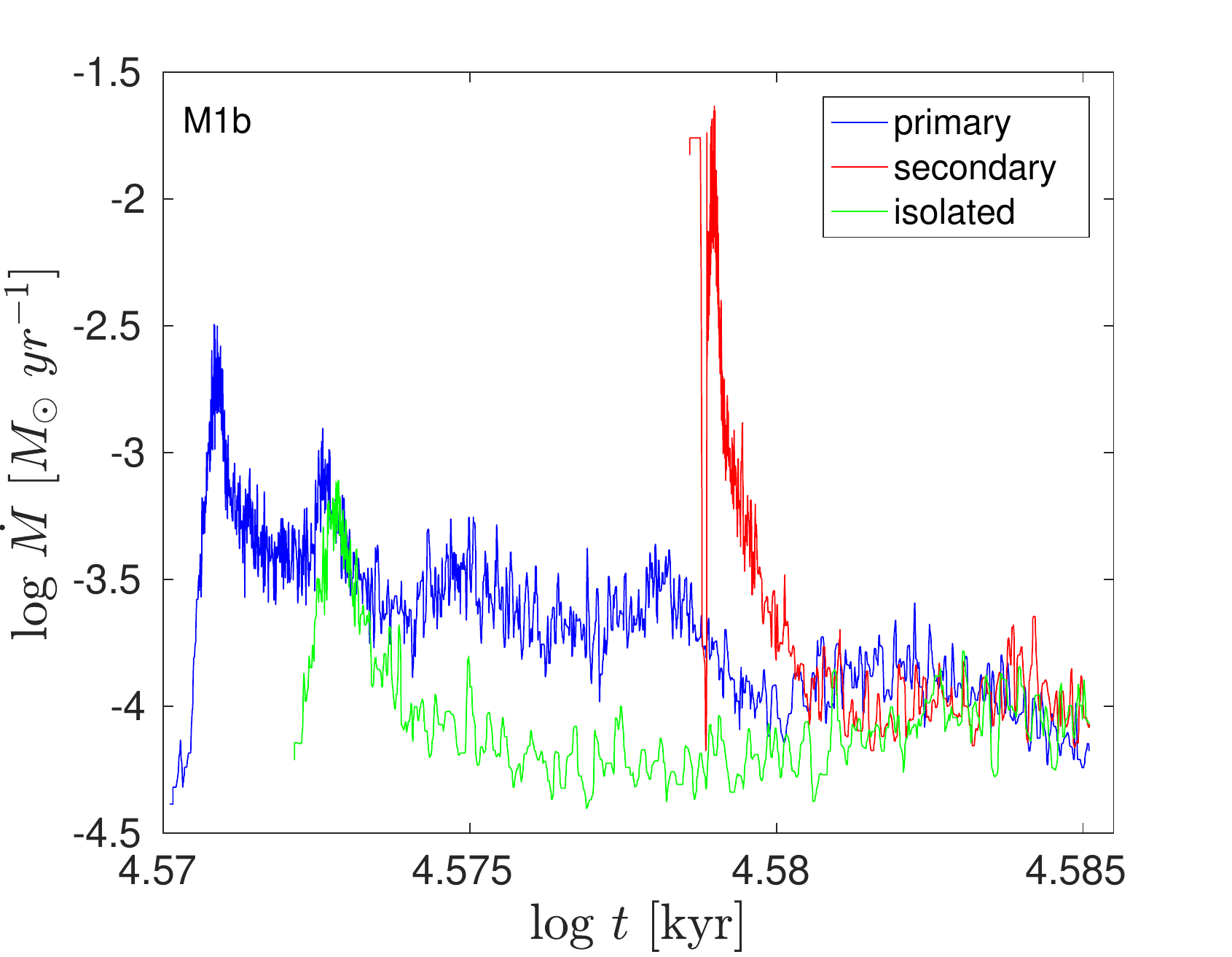}
	\includegraphics[width=\columnwidth]{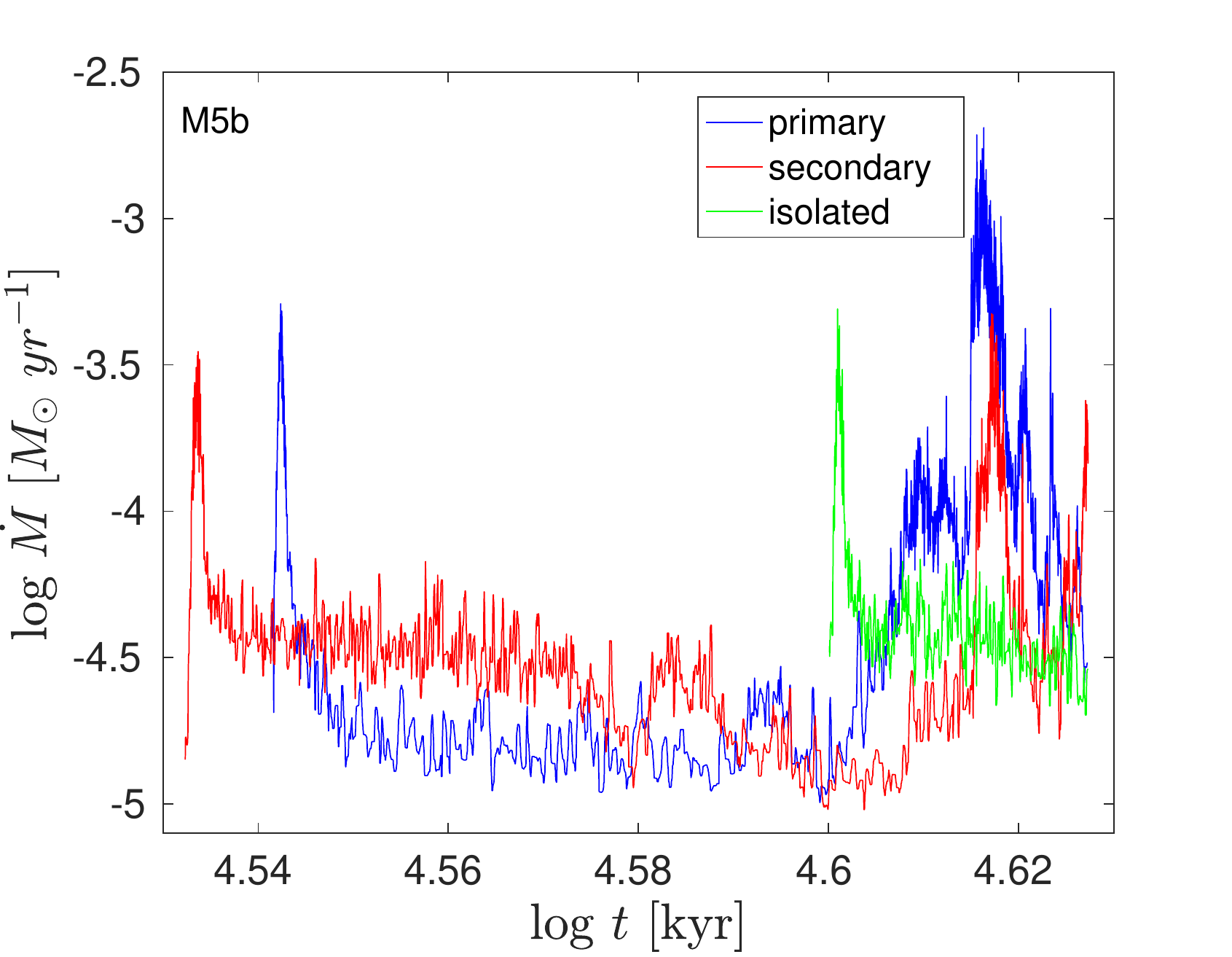}
	\caption{Accretion rates for the primary (blue), secondary (red), and isolated protostars (green) in models M1b (top panel) and M5b (bottom panel). The accretion rate is given in units of M$_{\odot}$ yr$^{-1}$ and the time is in kyr. Colour in online edition.}
	\label{fig:figure1}
\end{figure}

Figure 8, shows the $\dot M$ evolution of the most massive protobinary system and the most massive isolated protostar in models M1b and M5b in the top and bottom panels, respectively. In model M1b of the Kolmogorov-type turbulence (top panel), the primary component shown in blue forms at $t$ = 37.153 kyr while the secondary in red is created earlier at $t$ = 37.931 kyr. The first accretion bursts for the primary and secondary components of the binary system are exhibited with peak values of $\dot M_{\rm prim}$ = 1.4 $\times 10^{-3}$ M$_{\odot}$ yr$^{-1}$ at $t$ = 37.411 kyr, and  $\dot M_{\rm sec}$ = 2.3 $\times 10^{-2}$ M$_{\odot}$ yr$^{-1}$ at $t$ = 37.940 kyr, respectively. We see a general decreasing trend in $\dot M$ for both companions, especially for the secondary. The isolated protostar in this model appears at $t$ = 37.325 kyr and exhibits no signs of accretion burst event. During the remaining part of its evolution, the general decreasing trend in $\dot M$ continues down to $\dot M_{\rm iso}$ = 8.8 $\times 10^{-5}$ M$_{\odot}$ yr$^{-1}$ until we terminate the simulation at $t$ = 38.5 kyr.

\begin{figure}
	
	\centering
		\includegraphics[width=\columnwidth]{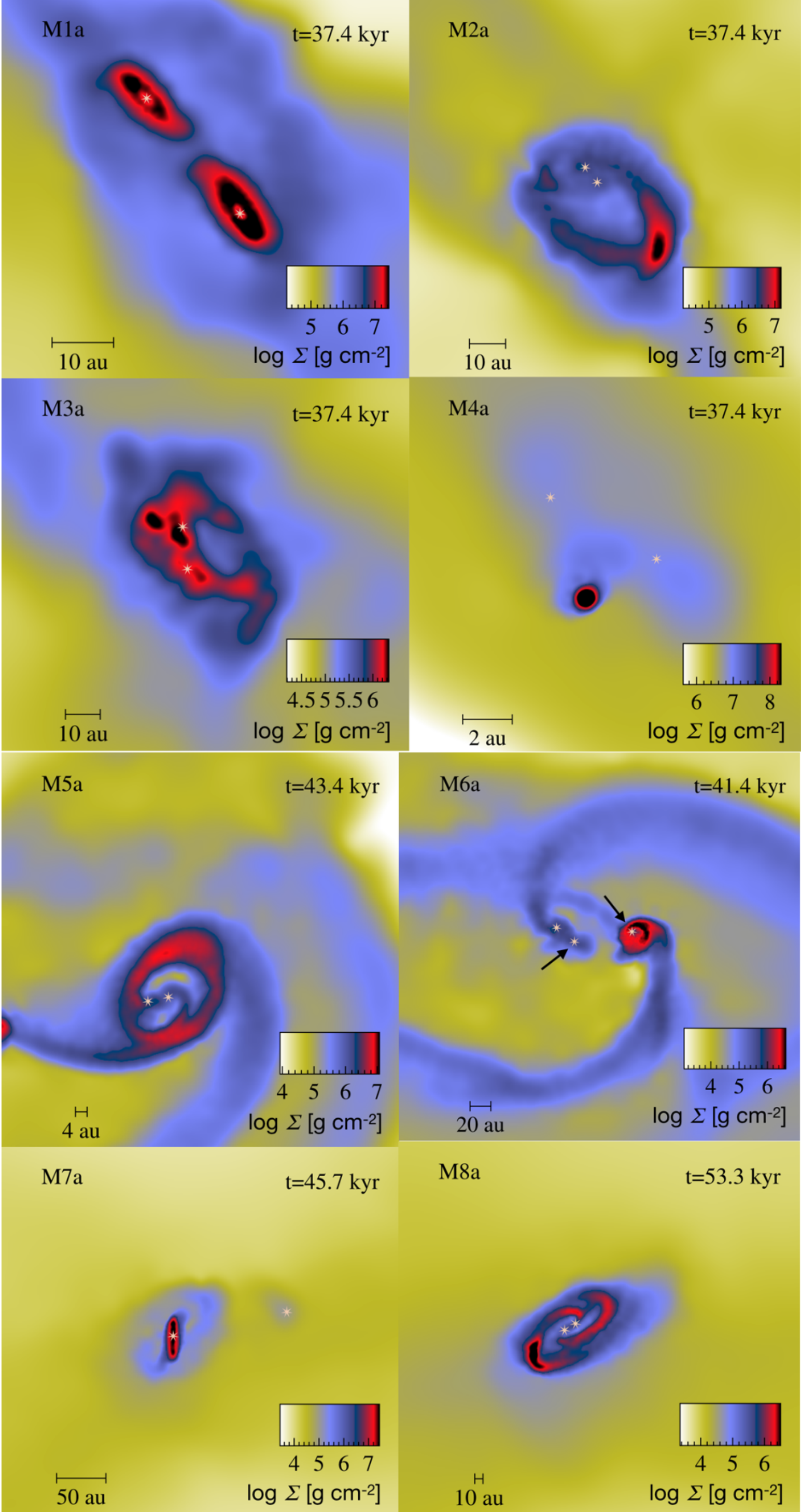}
	\caption{Simulation results of models M1a$-$M8a at their respective final states of evolution. Each panel shows the column density image in the xy-plane that focuses the protobinary system involved in exhibiting the accretion burst(s). The horizontal shaded bar inside each panel shows log ($\rm \Sigma{}$) in g cm$^{-2}$. The corresponding dynamical time in kyr is shown at the top-right corner of each panel. Colour in online edition.}
	\label{fig:figure1}
\end{figure}

In model M5b with a Burgers-type turbulence (bottom panel), the primary component shown in blue forms at $t$ = 34.833 kyr while the secondary in red is created at $t$ = 34.040 kyr. 
Until $t \approx$ 39.810 kyr, the protostars show an irregular low $\dot M$ evolution with less than a half order of magnitude variations. However, beyond $t \approx$ 39.810 kyr till the very end of the simulation, both protobinary companions exhibit vigorous growth in $\dot M$ with a series of intense accretion bursts. These strong accretion activity for both the primary and secondary companions shows peaks as high as $\dot M_{\rm prim}$ = 2.2 $\times 10^{-2}$ M$_{\odot}$ yr$^{-1}$ and $\dot M_{\rm sec}$ = 3.9 $\times 10^{-4}$ M$_{\odot}$ yr$^{-1}$, respectively. The variations in magnitude of $\dot M$ occasionally becomes even greater than an order of magnitude. The primary companion as compared to the secondary shows more intense accretion burst events of intervals even lesser than 1 kyr. This active accretion in primary component is an indicative of the surrounding material of lower specific angular momentum is falling into the binary system close to its centre of mass hence enabling the primary to accrete more than the secondary, as previously discussed by \citet{bate2000predicting,batel1997accretion,artymowicz1983role}. During the later phase of evolution, the two binary components continue to show episodes of intense mass accretion bursts. The isolated most massive protostar in this model originally appears at $t$ = 39.810 kyr. Since its formation, it shows insignificant mass accretion activity. However, the isolated protostar maintains a low $\dot M$ evolution with an average $\dot M_{\rm iso}$ = 2.8 $\times 10^{-5}$ M$_{\odot}$ yr$^{-1}$ until we terminate the simulation at $t$ = 42.4 kyr.

Figure 9 serves well to relate the mass accretion rates of the protobinary companions to the available gas in the surroundings. Channeled streams of gas provide a material supply to these individual companions as well as the protobinary system as a whole. The figure shows the morphology of the systems at the end of the simulations in eight panels each representing a specific model. The panel for model M1a shows the presence of a dense circumstellar disc around both the primary and secondary companions constituting the binary system. The existence of such a disc is an indicator that the rising trend in $\dot M$ at the end of the simulation (see Figure 7, top-panel) is due to the material which is efficiently transferred to the protostars from their individual circumstellar discs and not due to the Roche-lobe flow (we expect for the Roche-lobe flow that the material gain of one companion should result in a mass loss of the other component) \citep{gomez2016pre,hanawa2008accretion}. We, however, do not resolve individual stellar components and cannot simulate mass transfer in binaries. 

The panel for model M2a shows that the two companions of the protobinary system do not have well-defined individual circumstellar discs. The rising trend in the $\dot M$ evolution at the end of the simulation (see Figure A1, top-panel) is a result of the material-supply from the nearby dense gas structure which feeds the two protostars. The panel for model M3a indicates the two companions of the protobinary system that reside in the dense gas structure. The continuous phase of frequent accretion bursts in the later part of the model evolution (see Figure A2, top-panel) is most likely  the result of the ample supply of material from the nearby dense gas structure close to the protobinary system. Because the two companions exhibit mutually consistent episodes in the $\dot M$ evolution, a material flow via the Roche-lobe may be ruled out. The panel for model M4a also shows the absence of a well defined individual circumstellar disc structure around the two protostars. However, the presence of nearby dense gas structure provides enough material to the protobinary system. This results in mutually consistent accretion bursts observed in the  evolution of $\dot M$ for the two protostars (see Figure A3, top-panel). 

The panel for model M5a shows the presence of a strong circumbinary disc structure which is also connected to the nearby trail of spiralling dense gas. This has been observed and also remained as a feature of numerical studies where core collapse is either observed or modeled \citep{matsumoto2019structure,tang2017orbital,ohashi2014formation,artymowicz1996mass}. The intense episodes of mass accretion seen at the final stages of the model evolution (see Figure 7, bottom-panel) are most likely  the outcome of the gas inflow from the circumbinay disc onto the individual protostars. The panel for model M6a shows the presence of a strong circumstellar disc structure associated with one of the protobinary companions while the other protostar resides in the dense filamentary structure of the gas. A sharp rise in the $\dot M$ evolution of the primary companion (see Figure 8, bottom-panel) seems to be the result of the gas flow directly from the circumstellar disc onto the protostar. The less efficient accretion burst activity in the secondary companion could be the result of a less effective material supply via the lower-density filamentary structure the secondary companion resides in. The panel for model M7a shows the presence of a strong circumstellar disc structure associated with the primary companion while the secondary dwells in a much less dense structure of the gas. The strong and well defined circumstellar disc of the primary helps the protostar to maintain its $\dot M$ evolution which never shows a decline throughout the model evolution (see Figure A2, bottom-panel). Contrary to this, the  evolution of $\dot M$ of the secondary companion due to the inefficient material supply is continuously on a decline and never recovers from the unavailability of the surrounding gas. The panel for model M8a is a manifestation of the protobinary system that exists inside the dense circumbinary disc. The trails of gas flowing towards the two companions keep providing enough material for the two protostars to exhibit series of frequent accretion bursts that continue until the very end of the model evolution (see Figure 10, bottom-panel).    

Figure 10 serves the same purpose as Figure 9 but for the other random seed. The panel for model M1b shows the presence of a dense circumstellar disc around both the protostars which constitute the protobinary system. Surprisingly, the two companions do not exhibit any significant activity in their mass accretion rate evolution (see Figure 8, top-panel). The panel for model M2b shows the presence of dense circumstellar discs around both of the protostars. These individual circumstellar discs not only show a mutual gas flow but also remain connected with the external and more dense spiral gas structure. The frequent but mutually inconsistent accretion bursts seen in the two protostars (see Figure A4, top-panel) may well be the result of frequent gas inflow from the circumstellar discs onto the protostars. We suspect that the possible presence of Roche-lobe overflow regularly transfers material from one circumstellar disc to another giving birth to the mutually inconsistent series of accretion bursts of the two protostars. However, confirming this claim would require a higher spatial resolution. 

The panel for model M3b shows the presence of a dense and much wider circumbinary disc structure around the primary companion. The secondary companion, however, resides in the less dense gas structure but remains connected with the circumstellar disc of the primary. Despite the dense surroundings, the disc structure of the primary seems to remain isolated and disconnected from its surrounding gas reservoir. This can be the result of the general declining trend observed in the $\dot M$ evolution for the primary companion (see Figure A5, top-panel). The occasional accretion bursts seen for the secondary companion are most likely  due to its presence inside the dense gas structure that remains connected with the primary companion. 

The panel for model M4b shows the presence of a dense circumstellar disc structure around the primary companion. The secondary companion, however, resides in a less dense gas structure. The primary maintains a high $\dot M$ evolution with occasional boosts in its magnitude (see Figure A6, top-panel). The secondary despite the lower $\dot M$ evolution exhibits a few intense accretion bursts which are most likely  the result of the surrounding gas, which is less dense when compared with the gas in the proximity of the primary. The panel for model M5b indicates a dense circumstellar disc around the primary companion. A long dense trail of gas remains associated with the structure that provides the material infall towards the circumstellar disc of the primary. This is most likely the prime reason behind a series of intense accretion bursts exhibited by the primary companion (see Figure 8, bottom-panel) during the later stages of model evolution. The secondary companion, however, resides in a less dense gas structure which still remains connected to the more extended filamentary gas trail. This provides the secondary companion enough gas infall to exhibit a couple of intense accretion bursts at the later stages of the model evolution. 

\begin{figure}
	
	\centering
    \includegraphics[width=\columnwidth]{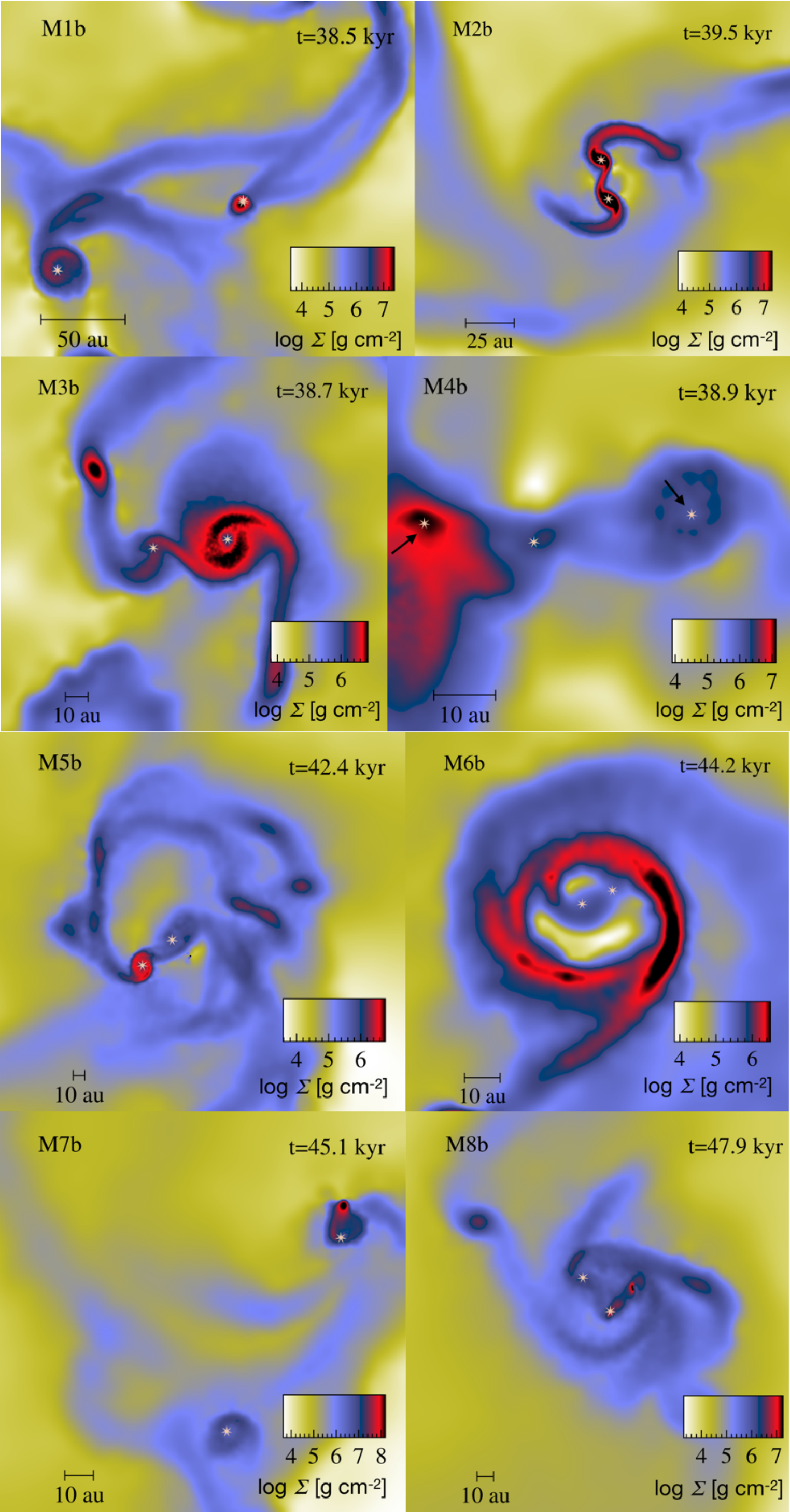}
	\caption{Simulation results of models M1b$-$M8b at their respective final states of evolution. Each panel shows the column density image in the xy-plane that focuses the protobinary system involved in exhibiting the accretion burst(s). The horizontal shaded bar inside each panel shows log ($\rm \Sigma{}$) in g cm$^{-2}$. The corresponding dynamical time in kyr is shown at the top-right corner of each panel. Colour in online edition.}
	\label{fig:figure1}
\end{figure}

The panel for model M6b shows the presence of gas concentrated in the circumbinary structure. The primary and secondary companions both lack the well defined individual circumstellar discs. This subsequently affects the  $\dot M$ evolution of the two companions which exhibit, in general, a low mass accretion rate at the end of the simulation (see Figure A4, bottom-panel). The panel for model M7b shows a more dense gas clump near the primary companion. The secondary companion remains far but yet connected via a dense trail of gas. The intense accretion bursts occur at the final moments of the model evolution suggesting  material flow from the nearby dense clump onto the primary companion (see Figure A5, bottom-panel). A less active $\dot M$ evolution which only shows a gradual rise in  $\dot M$ at the end of the simulation is mainly due to the lack of ample supply of material from the first to the circumstellar disc and then onto the protostar. The panel for model M8b shows the presence of a dense trail of spiralling gas which surrounds the protobinary system. At the later part of the model evolution, the $\dot M$ evolution for the two companions indicates a few accretion bursts (see Figure A6, bottom-panel). These accretion spikes are most likely linked to the trails of dense gas which provide material supply onto the two protostars.

\subsection{Binary properties ($m_{1}$ $m_{2}$, $d$, $a$, $e$, $q$ )}
 The binary systems appears as the most likely outcome of the collapsing gas cores \citep{riaz2020stellar,eggleton2008catalogue,klessen2000formation,boss1979fragmentation}. This is also true for our simulations. We present in Table 3 the characteristics of the most massive protobinary systems such as the masses of the two companions ($m_{1}$ $m_{2}$), the binary separation $d$, the semi-major axis $a$, the eccentricity $e$, and the mass ratio $q$. To investigate the possible correlation between these binary properties with the strongest accretion burst event(s) (symbolized as $\dot M_{\rm burst}$), we focus on the spike  observed in the $\dot M$ evolution for the two companions in each model.
 
 The data for the semi-major axis $a$ and mass ratio $q$ is obtained at the final stage of each model. This does not relate directly to the exact time of occurrence of maximum accretion burst. However, we notice that the strongest accretion events in our models occur close to the time when we terminate our models. We, therefore, expect that the estimated model parameters should be reasonably close to the ones at the moment of the peak, but are mostly meant to be indicative of the system. 
 
 \begin{figure}
	
	\centering
	\includegraphics[width=\columnwidth]{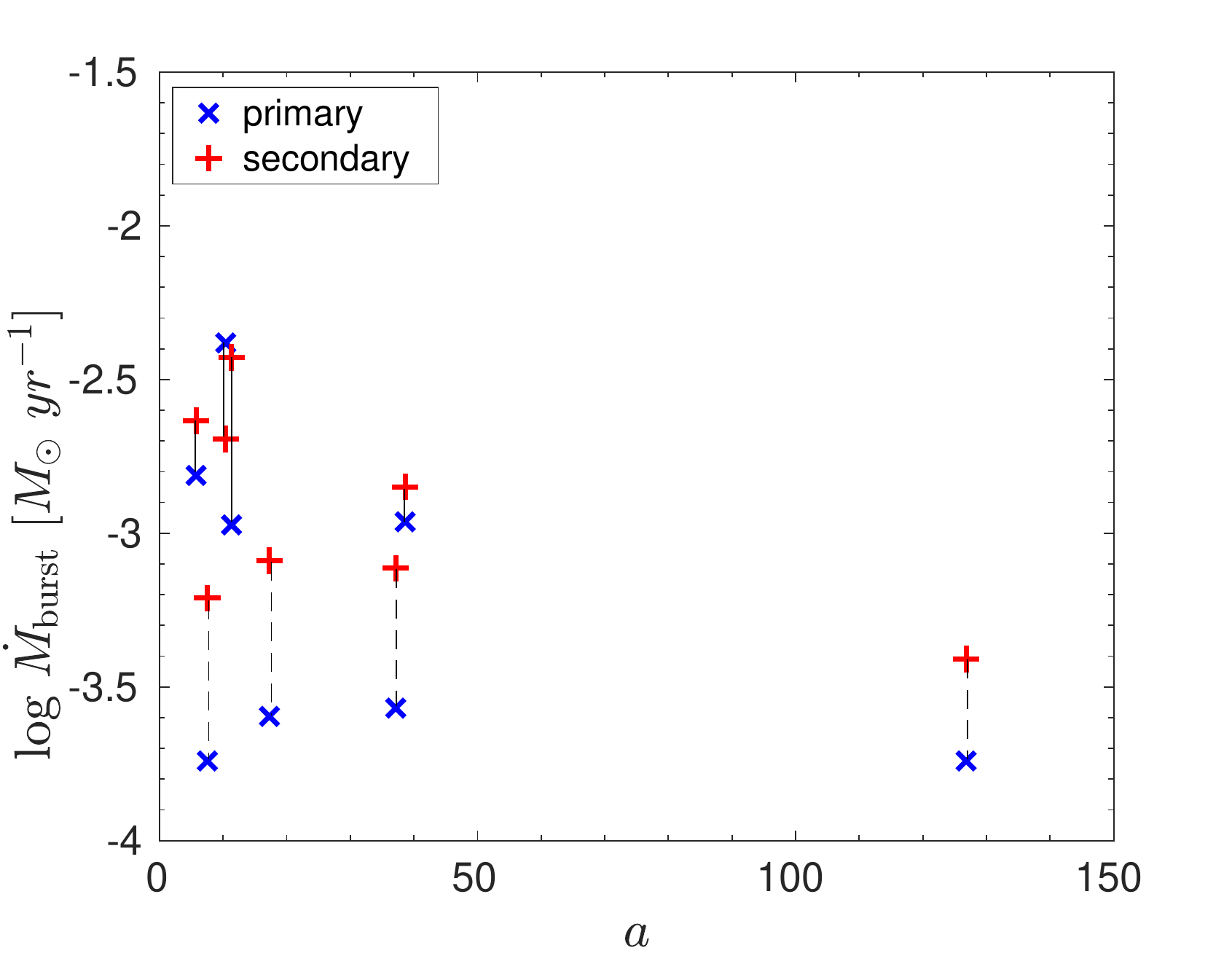}
	\includegraphics[width=\columnwidth]{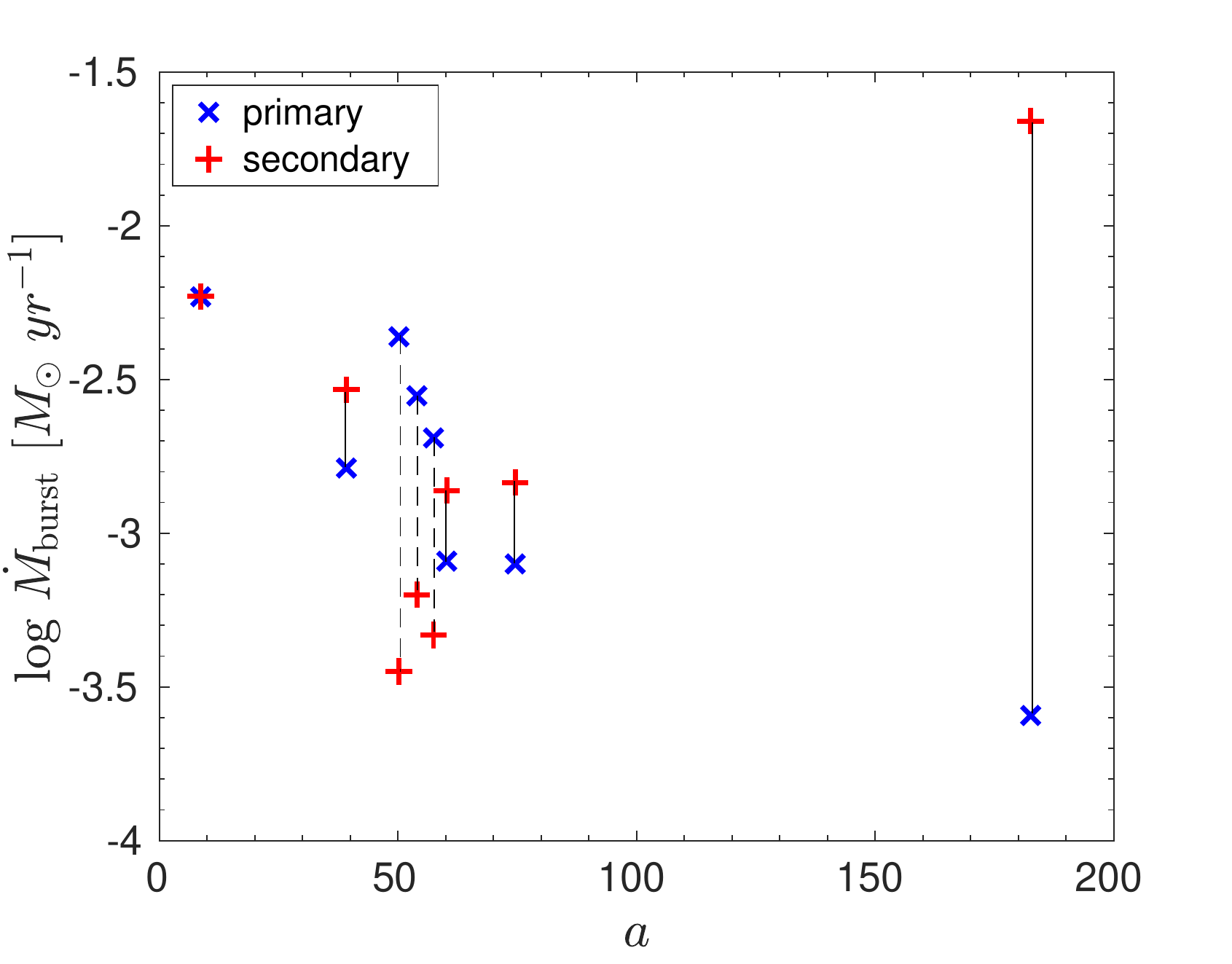}

	\caption{The strength of accretion bursts of the protobinary systems as a function of its semi-major axis. The top and bottom panels are for the models M1a$-$M8a and M1b$-$M8b, respectively. The blue cross and the red plus markers are for the primary and the secondary companions of the protobinary system, respectively. The vertical solid and dashed lines connecting the two components of the most massive binary system represent models that follow Kolmogorov and Burgers-type turbulence models, respectively. The accretion burst is given in units M$_{\odot}$ yr$^{-1}$ and the semi-major axis in au. Colour in online edition.}
	\label{fig:figure1}
\end{figure}
 
 In Figures 11 and 12, the top and bottom panels in each figure show the results of the models based for seeds 1 and 2, respectively. The blue cross and red plus markers represent the properties related to the primary and the secondary companion of the protobinary system, respectively. 
  Figure 11 (top and bottom panels) illustrates the strongest mass accretion burst $\dot M_{\rm burst}$ from the protobinary system as a function of the semi-major axis $a$. The protobinary systems of small semi-major axis $a$ $\approx$ 50 au support intense accretion burst.
  Figure 12 (top and bottom panels) shows the strongest mass accretion burst $\dot M_{\rm burst}$ from the protobinary system as a function of the mass ratio $q$. The protobinary system with high mass ratio  $q$ > 0.7 seems to support the accretion burst event in the secondary companion, regardless of the nature of turbulence prevailing in the collapsing gas core. However, it is observed in the second seed case that the primary component of the binary systems of $q$ $\leq$ 0.51 remains more active in exhibiting strong accretion bursts. \citet{girichidis2012importance} and  \citet{stacy2012first} have discussed that the mass ratio $q$ of the system plays an important role to determine which component of the binary system will accrete more from the channelled stream of the surrounding gas. Also, \citet{bate1997accretion} have shown that if the surrounding material of low specific angular momentum (AM) falls into the binary system then it most likely falls at the centre of mass (CM) of the binary system. This facilitates the primary to accrete this infalling material more easily and subsequently to produce strong $\dot M_{\rm burst}$. However, an infall of gas with high specific AM supports the secondary companion to accrete more and to show a strong $\dot M_{\rm burst}$ as the gas then falls into its periastron distance from the CM of the system. We notice that the cases in which the primary exhibits more intense accretion burst than the secondary occur for Burgers-type turbulence, provided that the system has a low mass ratio (see Figure 12, bottom-panel). Whereas, the opposite is found in gas cores that have Kolomogorov-type turbulence where binary systems have a high mass ratio (see Figure 12, top-panel). We therefore suspect that the material infall from the immediate surroundings into the binary systems that reside in gas cores of Burgers-type turbulence must be of low specific AM (at least, for the cases where binary systems have $q$ $\leq$ 0.51). In case of gas cores of Kolmogorov-type turbulence, which host the binary system there must be a high specific AM meterial infall into the binary systems that have $q$ > 0.7. These binary systems must be on the path of dynamical evolution to eventually become equal mass binaries. 
  
  \begin{figure}
	
	\centering
	\includegraphics[width=\columnwidth]{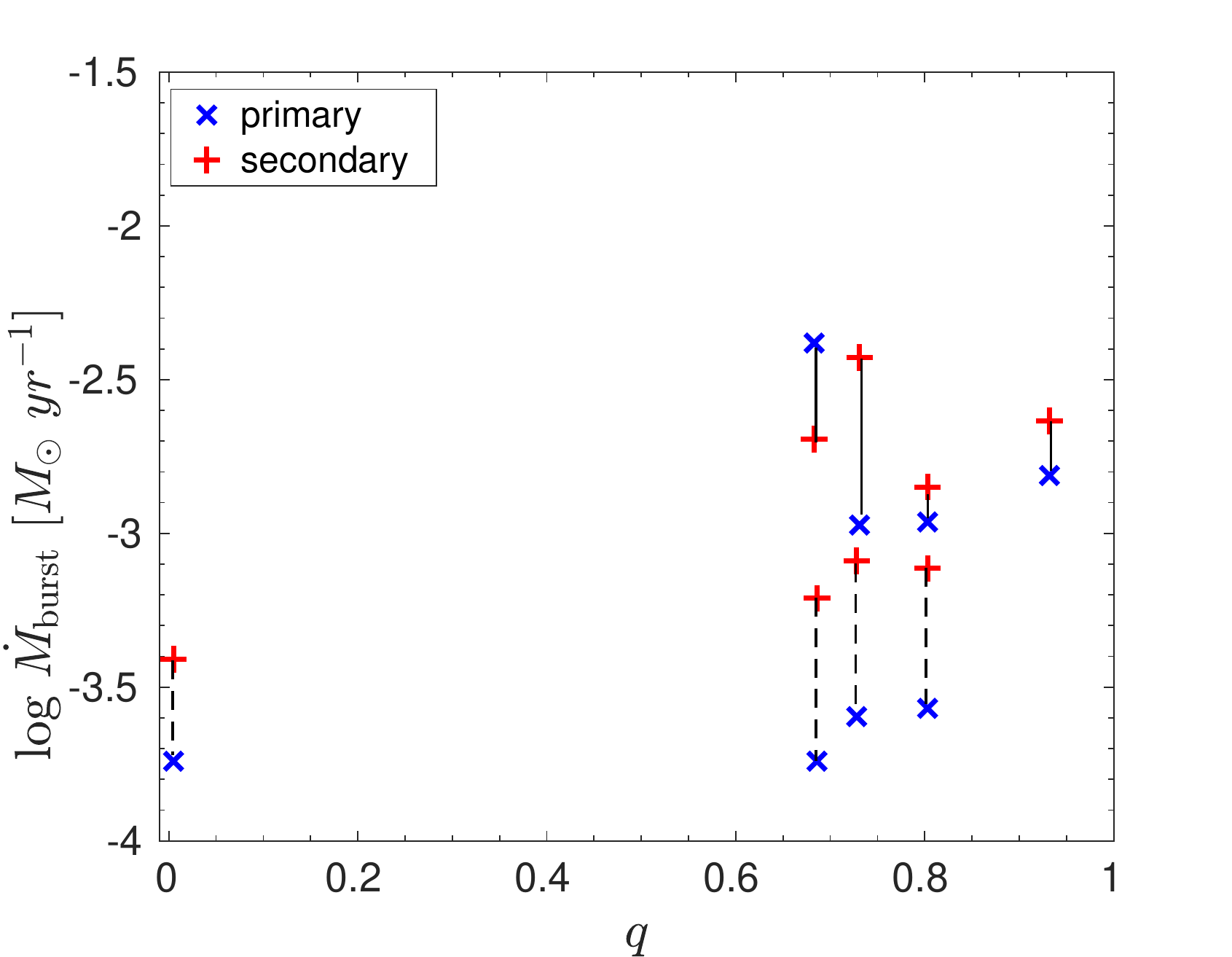}
	\includegraphics[width=\columnwidth]{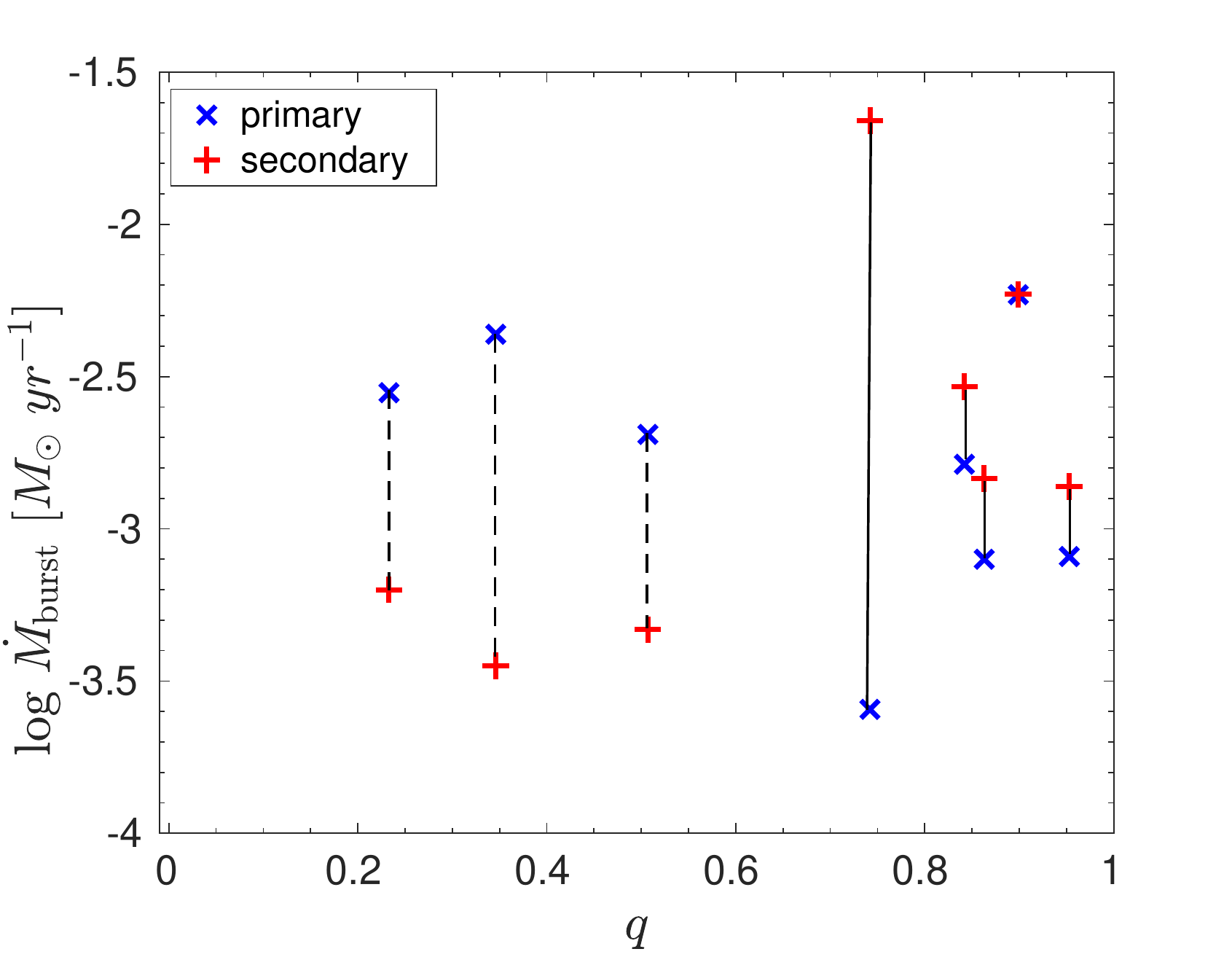}

	\caption{The strength of accretion bursts of the protobinary system as a function of its mass ratio. The top and bottom panels are for the models M1a$-$M8a and M1b$-$M8b, respectively. The blue cross and the red plus markers are for the primary and the secondary companions of the protobinary system, respectively. The vertical solid and dashed lines connecting the two components of the most massive binary system represent models that follow Kolmogorov and Burgers-type turbulence models, respectively. The accretion burst is given in units M$_{\odot}$ yr$^{-1}$. Colour in online edition.}
	\label{fig:figure1}
\end{figure}
  
  We also analysed the strongest mass accretion burst $\dot M_{\rm burst}$ in the context of the binary separation $d$ and the orbital eccentricity $e$ of the protobinary systems. However, we find a weaker dependence for the former and almost no dependence on the latter.   
  
\subsection{Caveats}
The numerical scheme we implement to model the collapse of gravoturbulent gas does not take into account magnetic field which can add further to the turbulent support that the gas feels against self-gravity. The magnetic field has the potential to also limit the number of fragments \citep{commerccon2011collapse,peters2011interplay,price2008effect}. However, \citet{wurster2019there} have suggested that the formation of a protostellar disc depends more on the turbulence in the gas than on the magnetic field. Nonetheless, a moderately strong magnetic field can also suppress disc formation in the ideal MHD limit \citep{li2011non}. The formation of sub-Keplerian discs also known as magnetic braking catastrophe is yet another important aspect that needs to be investigated
\citep{seifried2012magnetic, seifried2012disc} and \citep{hennebelle2011collapse,hennebelle2009disk}.
The other limitation of our numerical scheme is that we neglect radiative feedback processes, which may become relevant for episodic accretion \citep{mercer2016effect,stamatellos2012episodic} and for the IMF \citep{hennebelle2020role}. Also, our findings suggest that the accretion bursts on the binary components are most likely connected to the tidal forces from the other stars and the availability of the gas supply. We, however, do not have enough resolution in our simulations to quantify this aspect.

\section{Discussion and conclusions}
The gravoturbulent collapse of gas with Kolmogorov-type turbulence leads to the formation of spiral arms and sharp density contrasts in the gas. However, in case of Burgers-type turbulence the collapsing gas produces density enhancements that appear as more diffused structures. This has consequences for the nature of fragmentation and the associated strength of the episodic accretion in protostars. We find that in most of the simulations reported here, the former produces nearly an order of magnitude more intense accretion bursts than the latter.   

Molecular gas cores with Kolmogorov-type turbulence with lower initial thermal states are found to form a larger number of protostars than the warmer gas cores. However, gas cores of Burgers-type turbulence in general produce relatively fewer numbers of protostars regardless of their initial thermal states. In most of the cases, the protobinary systems that emerge in the Kolmogorov case remain surrounded by dense structures in the form of long trails of gas. This leads to gas infall onto the protobinary stars causing them to exhibit episodes of intense accretion. 

Binary properties such as the semi-major axis and the mass ratio show trends in which the presence of the strongest accretion burst relates to the binary systems with $a$ $\approx$ 50 au. Similarly, systems with $q$ > 0.7 also seem to support the phenomenon of episodic accretion in gravoturbulent gas cores. Moreover, we find hints that the primary companion remains more active in showing the strongest mass accretion burst $\dot M_{\rm burst}$ when the binary system resides in a  core with Burgers-type turbulence and has $q$ $\leq$ 0.51, while exactly the opposite is observed in most of the gas cores where Kolomogorov-type turbulence accompanies the gravitational collapse of the gas. This may well be an indication that the former has a low specific AM gas falling into the binary system from its immediate surroundings while the latter has a material infall of high specific AM from its nearby regions. We do find some of the binaries in the environments of Burgers-type turbulence where the secondary companion still remains more active, however, these system are of higher $q$. The cases of low mass ratio $q$ $\leq$ 0.51, found in the same type of turbulence support the binaries to evolve into systems of extreme mass ratios. Protobinaries with $a$ $\leq$ 50 au support the phenomenon of episodic accretion in gravoturbulent gas cores. Also, protobinary systems with $q$ > 0.7 show the phases of episodic accretion. We do not find any strong dependence on eccentricity, where we investigated a range from $e$ = 0 $-$ 1, and the protostellar response in terms of intense accretion bursts appears independent of the shape of the orbit.

 \citet{forgan2010stellar} have reported the secondary companion of the binary system as a more active companion in the protostellar systems in terms of intense accretion bursts when the influence of stellar encounters on disc dynamics was investigated for both Kolmorogov-like and Burgers-like models. We present a plausible scenario in which there can be protobinary systems in models of Burgers-type turbulence where the primary is more actively accreting companion than secondary. \citet{vorobyov2006burst} have shown that the accretion bursts remain associated with the formation of dense protostellar/protoplanetary embryos, which are later driven onto the protostar by the gravitational torques that develop in the disc. It is expected from the deeply young embedded protostars that they can exhibit variations in their accretion rates which can be tracked indirectly by following the response of the dust envelope at mid-IR to millimeter wavelengths \citep{francis2019identifying}. We believe that such observations can be even more plausible for the secondary companion of the embedded protobinary system if the gas in the star-forming region is found with a subsonic velocity dispersion. \citet{contreras2019determining} have found that outbursts in the Class II stage have a duration of $\sim$112 kyr which is 10 times less frequent than during the Class I stage. The accretion burst recurrence time-scale that we found in our simulations primarily for the class 0 stage objects is of the order of 1 kyr. Our models have shown a stronger activity of accretion bursts for close protobinary systems with $a$ $\leq$ 50 au. This seems consistent with the effort made to understand the protostellar accretion histories towards individual sources by utilising sublimation and freeze-out chemistry of CO \citep{frimann2017protostellar}.








\section*{Acknowledgements}
This research was partially supported by the supercomputing infrastructure of the NLHPC (ECM$-$02). The authors acknowledge the Kultrun Astronomy Hybrid Cluster (projects ANID Programa de Astronomia Fondo Quimal QUIMAL 170001, ANID PIA ACT172033, and Fondecyt Iniciacion 11170268) for providing HPC resources that have contributed to the research results reported in this paper. RR and the second author DRGS thank for funding through Fondecyt Postdoctorado (project code 3190344). Also, the Geryon
cluster at the Centro de Astro-Ingenieria UC was extensively used for the 
calculations performed in this paper. BASAL CATA PFB-06, the Anillo ACT-86, 
FONDEQUIP AIC-57, and QUIMAL 130008 provided funding for several improvements 
to the Geryon cluster. DRGS further thanks for funding via Fondecyt regular (project code 1161247) and via the Chilean BASAL Centro de Excelencia en Astrof\'isica yTecnolog\'ias Afines (CATA) grant PFB-06/2007. SV wishes to thank Prof. Dr. R. Keppens and Prof.  Dr. S. Poedts for providing access to the KUL supercomputing cluster Thinking while developing and testing the code that was used in this work. He also gratefully acknowledges the support of the KUL HPC team. RSK acknowledges financial support from the German Research Foundation (DFG) via the collaborative research centre (SFB 881, Project-ID 138713538) ``The Milky Way System'' (subprojects B1, B2, and B8) and from the Heidelberg cluster of excellence EXC 2181 (Project-ID 390900948) ``STRUCTURES: A unifying approach to emergent phenomena in the physical world, mathematics, and complex data'' funded by the German Excellence Strategy.

\section*{Data Availability}
The data underlying this article will be shared on reasonable request to the corresponding author.


\bibliographystyle{mnras}
\bibliography{rsvk_accepted} 




\appendix

\section{Mass accretion rates}
In this section, we continue describing how the two companions of the most massive protobinary system along with the most massive isolated protstar (if exists) appearing in model sets M2a - M8a and M2b - M8b exhibit their evolution of mass accretion rates.

In Figure A1, the evolution of $\dot M$ is shown for the most massive protobinary system and the most massive isolated protostar (if present) in models M2a and M6a in the top and bottom panels, respectively. In model M2a with a Kolmogorov-type turbulence flows (top panel), the primary component shown in blue appears at $t$ = 36.438 kyr while the secondary in red forms at $t$ = 36.479 kyr. For a brief interval until $t$ $\approx$ 36.812 kyr the protobinary system remains less active in producing intense accretion bursts. Beyond this the two components exhibit the first accretion bursts with peak values of $\dot M_{\rm prim}$ = 31.1 $\times 10^{-3}$ M$_{\odot}$ yr$^{-1}$ and $\dot M_{\rm sec}$ = 3.0 $\times 10^{-3}$ M$_{\odot}$ yr$^{-1}$. The two components during the later phase of their evolution continue to exhibit intense and frequent accretion bursts until we terminate the simulation at $t$ = 37.4 kyr. The isolated massive protostar depicted in green is formed at $t$ = 37.294 kyr. There are no accretion bursts associated with the isolated protostars. However, during the course of its evolution until we terminate the simulation, it exhibits a gradually rising trend in the evolution of $\dot M$ and the peak attains $\dot M_{\rm iso}$ = 6.0 $\times 10^{-4}$ M$_{\odot}$ yr$^{-1}$. 
In model M6a with a Burgers-type turbulence (bottom panel), the primary component shown in blue forms at $t$ = 36.475 kyr while the secondary in red forms at t = 38.815 kyr. The interplay between the two components is likely to begin after $t$ $\approx$ 39.355 kyr. We find that the primary shows relatively less intense accretion bursts maintaining around half an order of magnitude fluctuations in $\dot M$ with a peak of $\dot M_{\rm prim}$ = 2.7 $\times 10^{-4}$ M$_{\odot}$ yr$^{-1}$ until $t$ $\approx$ 41.114 kyr. However, the secondary component exhibits less frequent but intense accretion bursts. After its formation, the most intense accretion burst comes at $t$ = 39.536 kyr and the peak reaches $\dot M_{\rm sec}$ = 7.5 $\times 10^{-4}$ M$_{\odot}$ yr$^{-1}$. Later on, the trend in $\dot M$ remains decreasing but the secondary companion keeps exhibiting random but intense accretion bursts. After $t$ $\approx$ 41.114 kyr the two companions start showing mutually consistent and more intense accretion burst of nearly an order of magnitude higher until we terminate the simulation at $t$ = 41.4 kyr. The isolated protostar shown in green appears at t = 39.264 kyr. It experiences no accretion burst event. During the rest of its evolution until $t$ = 41.4 kyr it only manages to sustain the $\dot M$ evolution while exhibiting less than half of an order of magnitude fluctuations.         
\begin{figure}
	
	\centering
	\includegraphics[width=\columnwidth]{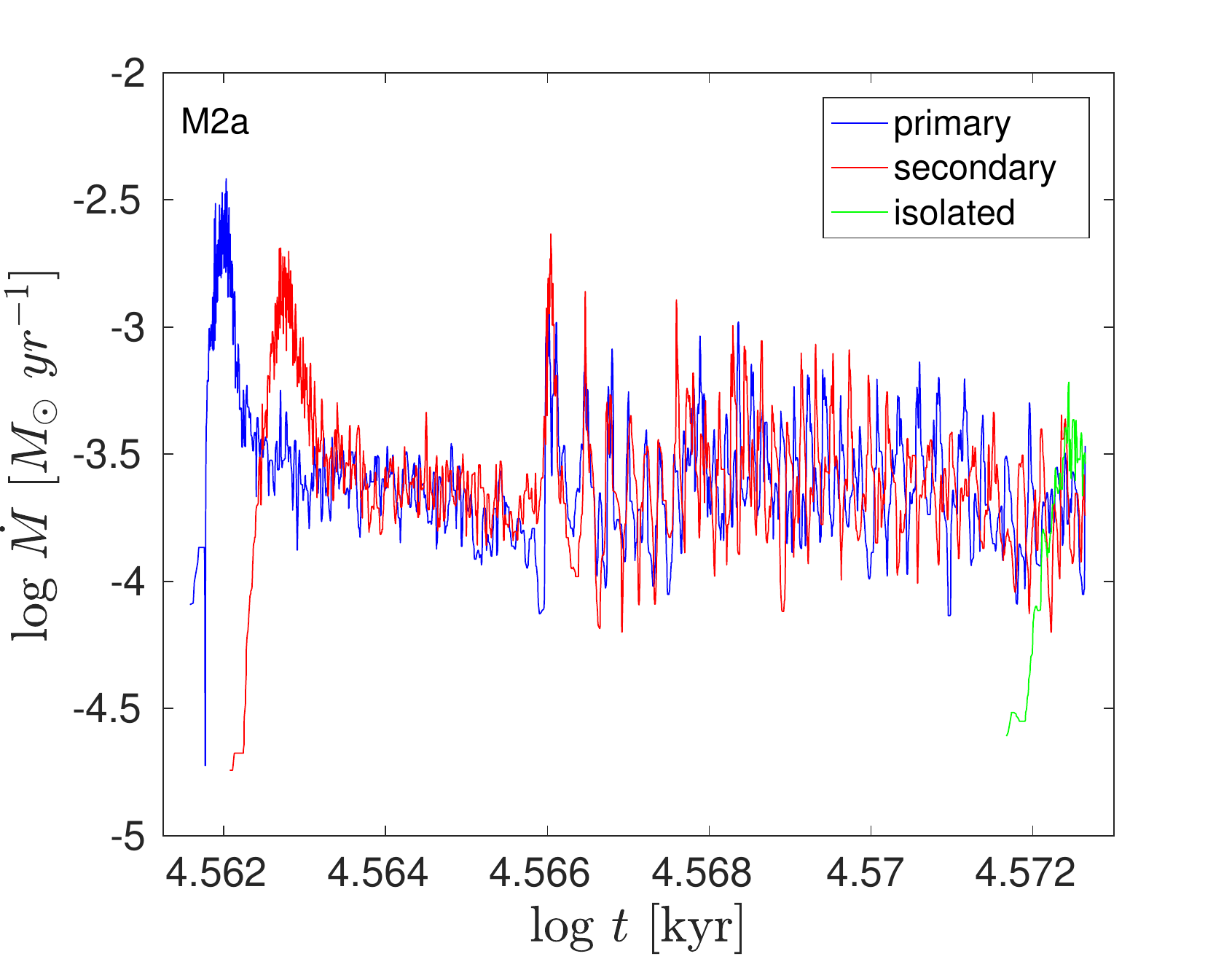}
	\includegraphics[width=\columnwidth]{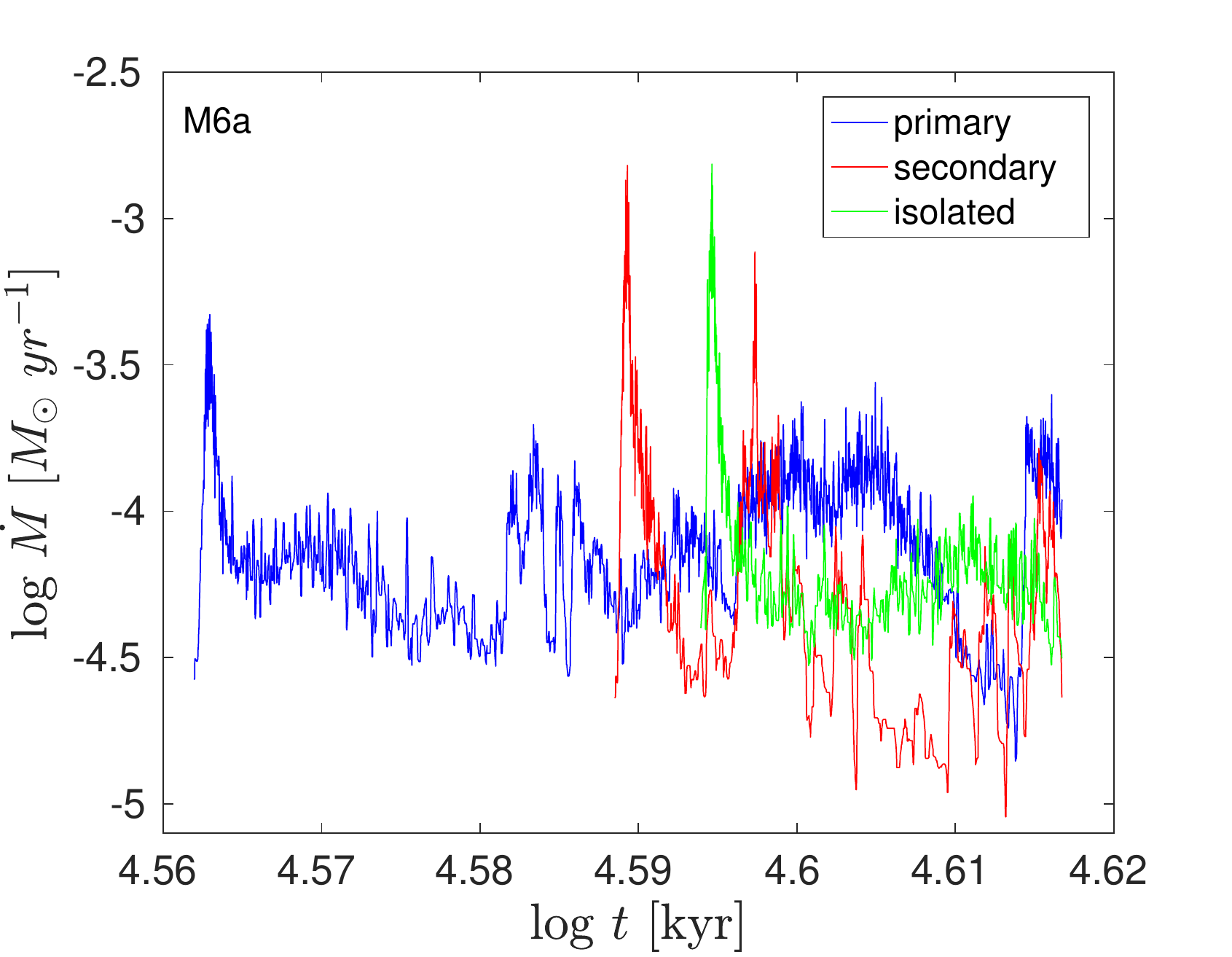}
	\caption{Accretion rates for the primary (blue), secondary (red), and isolated protostars (green) in models M2a (top panel) and M6a (bottom panel). The accretion rate is given in units of M$_{\odot}$ yr$^{-1}$ and the time is in kyr. Colour in online edition.}
	\label{fig:figure1}
\end{figure}

Figure A2, illustrates the $\dot M$ evolution of the most massive protobinary system and the most massive isolated protostar (if present) in models M3a and M7a in the top and bottom panels, respectively. In model M3a with a Kolmogorov-type turbulence (top panel), the primary component shown in blue arrives at $t$ = 36.475 kyr while the secondary in red is created at $t$ = 36.391 kyr. Soon after their formation the two companions during the course of their evolution show peaks in $\dot M$. The primary exhibits $\dot M_{\rm prim}$ = 1.3 $\times 10^{-3}$ M$_{\odot}$ yr$^{-1}$, while the secondary shows $\dot M_{\rm sec}$ = 1.5 $\times 10^{-3}$ M$_{\odot}$ yr$^{-1}$. After 36.643 kyr, the trend in $\dot M$ for both companions continues to decline, for the secondary in particular. However, from t $\approx$ 36.982 kyr onward, both the primary and secondary of the protobinary system start exhibiting a series of mutually consistent accretion bursts with a frequency of around 1 kyr. During this phase of evolution that lasts until we terminate the simulation at $t$ = 37.4 kyr, the first accretion burst associated with the secondary companion turns out to be more than one and a half order of magnitude large with a peak of $\dot M_{\rm sec}$ = 3.7 $\times 10^{-3}$ M$_{\odot}$ yr$^{-1}$. The isolated protostar whose $\dot M$ evolution is shown in green appears at $t$ = 36.897 kyr. It shows a first accretion burst $\dot M_{\rm iso}$ = 1.2 $\times 10^{-3}$ M$_{\odot}$ yr$^{-1}$  at $t$ = 36.982 kyr. The remaining phase of its evolution, however, shows a declining trend in $\dot M$ and the mass accretion rate at the end drops down to 8.9 $\times 10^{-5}$ M$_{\odot}$ yr$^{-1}$.

In model M7a with the Burgers-type turbulence (bottom panel), the primary component shown in blue forms at $t$ = 43.651 kyr while the secondary in red is created at $t$ = 44.463 kyr. The primary companion in the protobinary system of this model shows no accretion burst event. During the model evolution it maintains a consistent average $\dot M$ evolution of $\dot M_{\rm prim}$ = 8.3 $\times 10^{-5}$ M$_{\odot}$ yr$^{-1}$. The secondary companion in red shows its first accretion bursts of $\dot M_{\rm sec}$ = 3.9 $\times 10^{-4}$ M$_{\odot}$ yr$^{-1}$ at $t$ = 44.560 kyr. After this, it exhibits a gradually decreasing trend in $\dot M$ during the rest of the model evolution. The $\dot M$ drops down to 2.2 $\times 10^{-6}$ M$_{\odot}$ yr$^{-1}$ until we terminate the simulation at $t$ = 45.7 kyr. Also, at the end of our simulation no isolated protostar exists in model M7a. 

\begin{figure}
	
	\centering
	\includegraphics[width=\columnwidth]{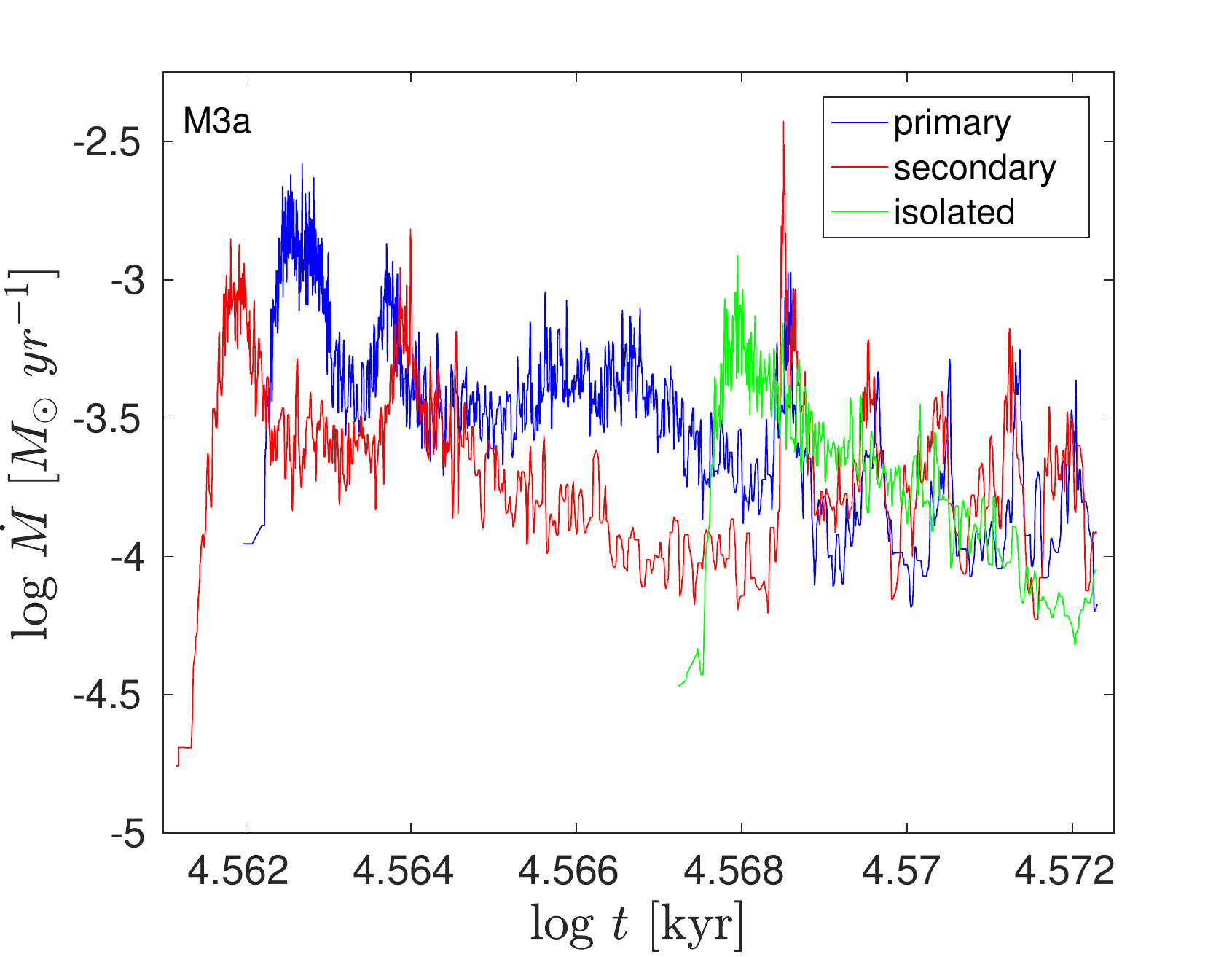}
	\includegraphics[width=\columnwidth]{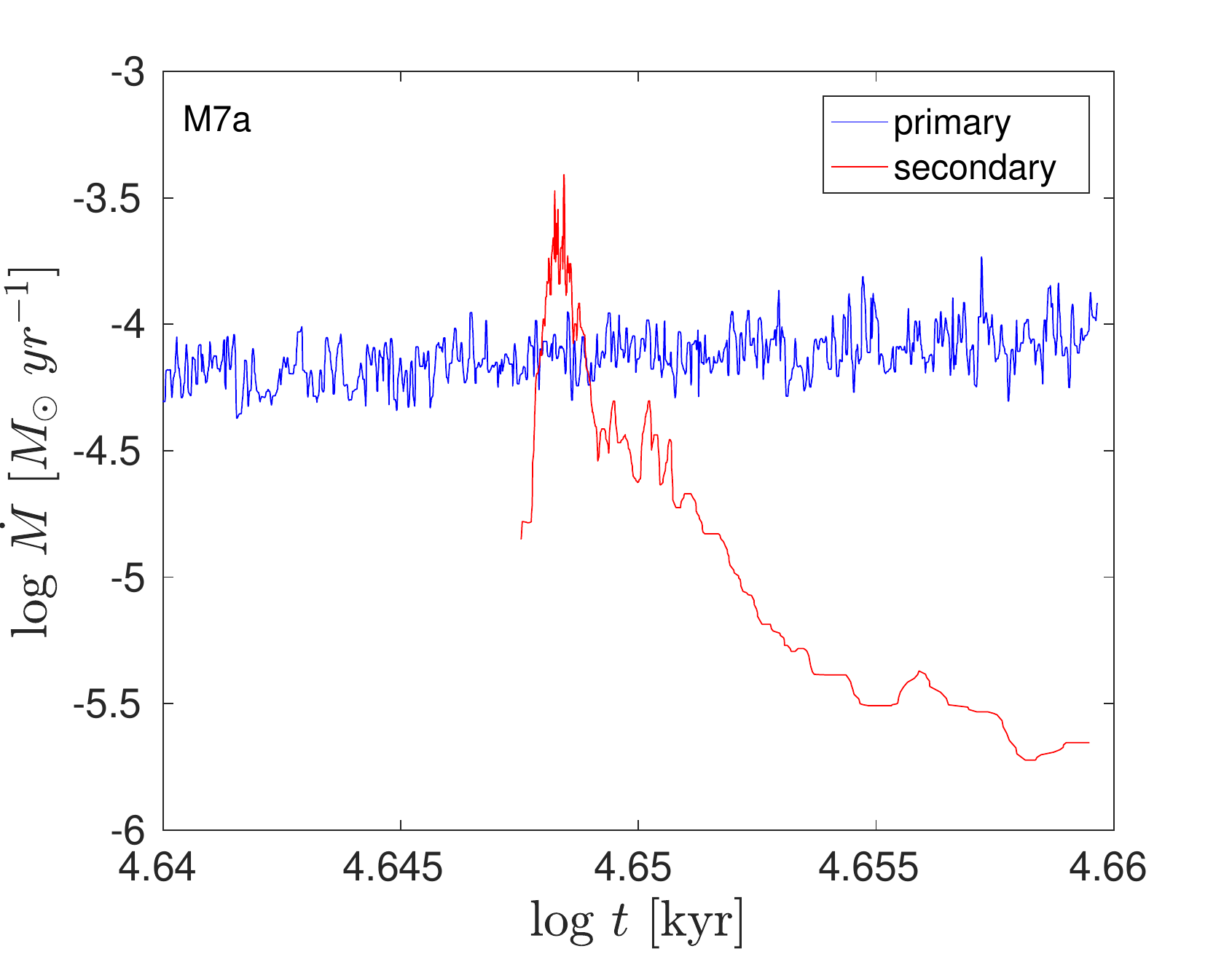}
	\caption{Accretion rates for the primary (blue), secondary (red), and isolated protostars (green) (if present) in models M3a (top panel) and M7a (bottom panel). The accretion rate is given in units of M$_{\odot}$ yr$^{-1}$ and the time is in kyr. Colour in online edition.}
	\label{fig:figure1}
\end{figure}
\begin{figure}
	
	\centering
	\includegraphics[width=\columnwidth]{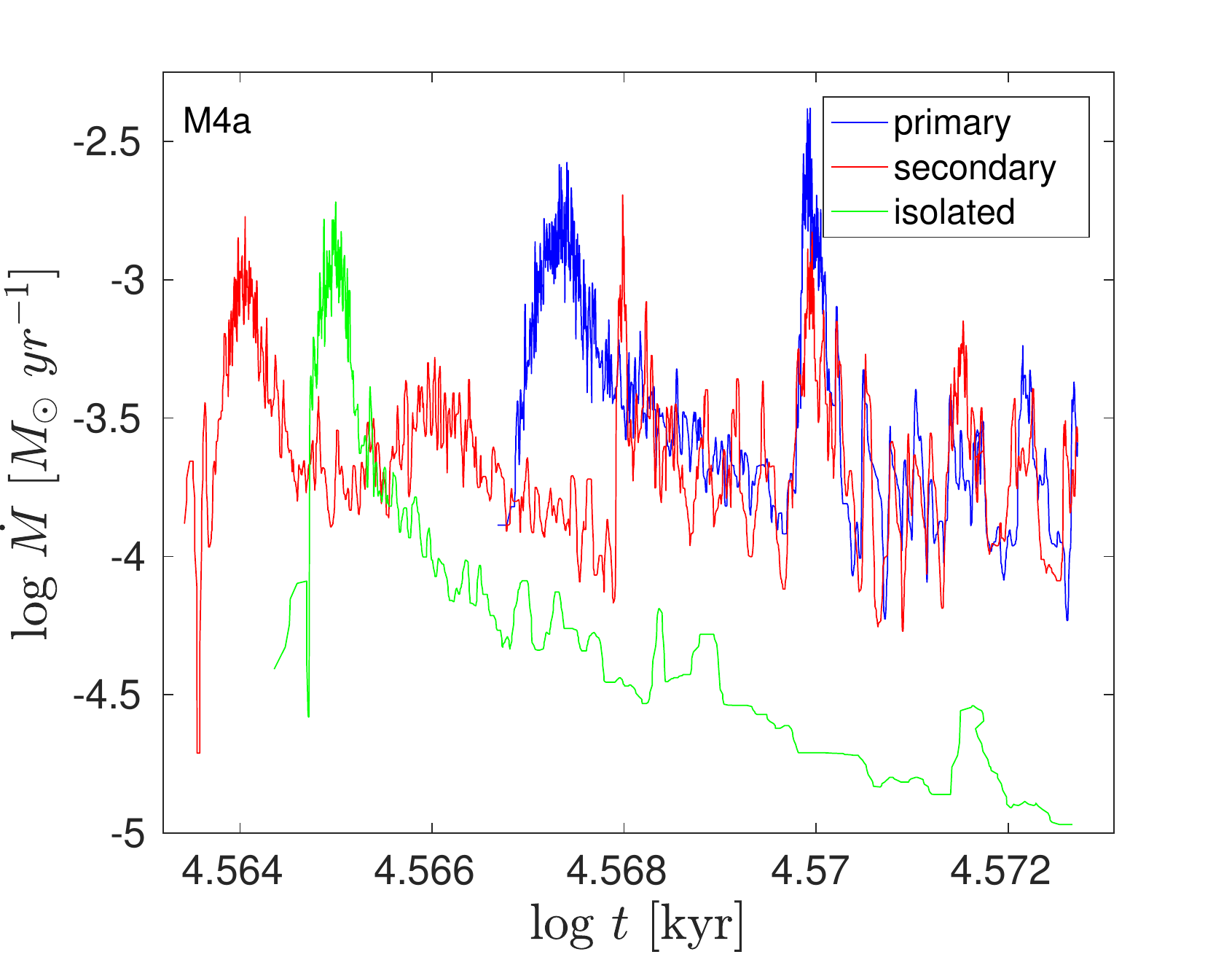}
	\includegraphics[width=\columnwidth]{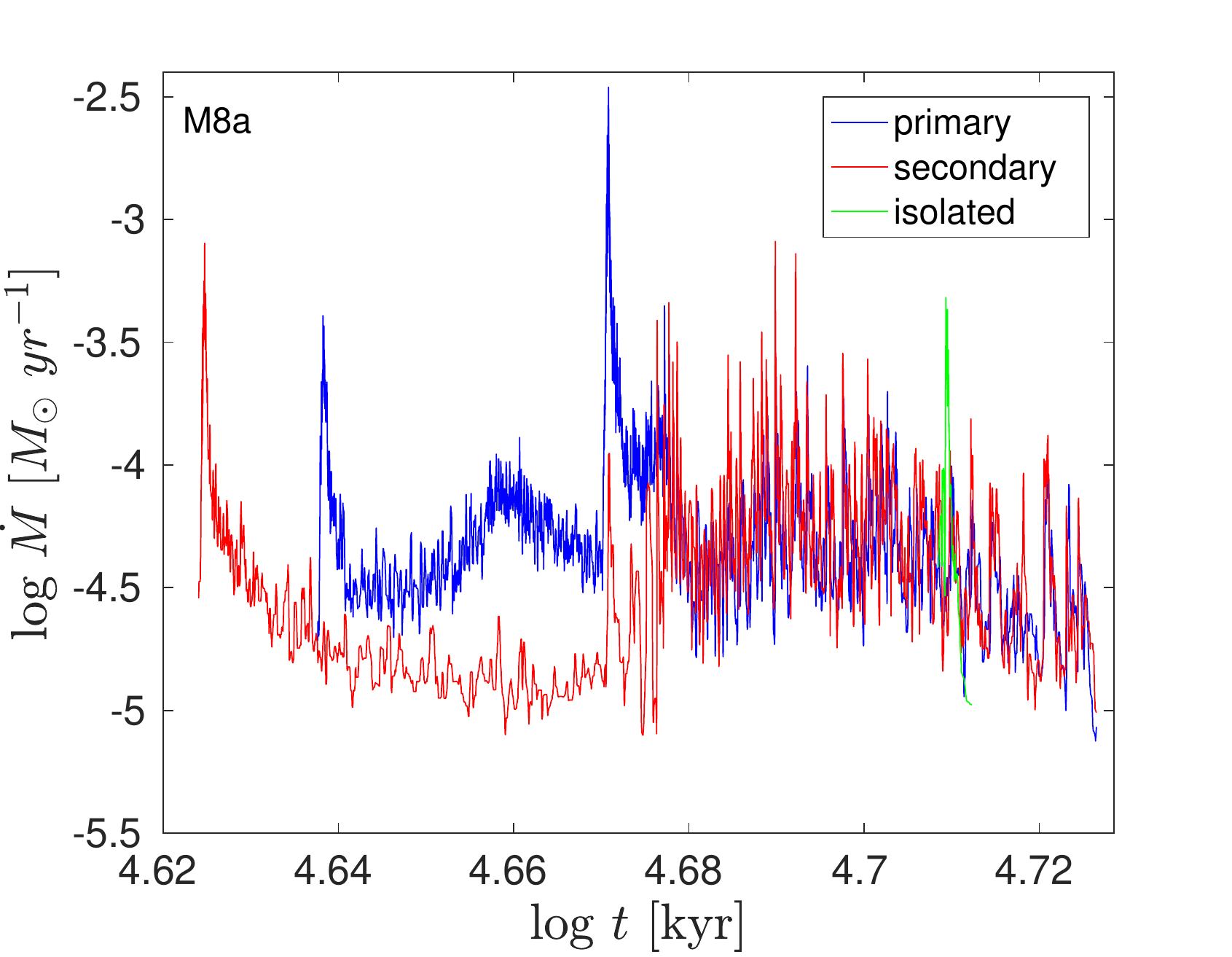}
	\caption{Accretion rates for the primary (blue), secondary (red), and isolated protostars (green) in models M4a (top panel) and M8a (bottom panel). The accretion rate is given in units of M$_{\odot}$ yr$^{-1}$ and the time is in kyr. Colour in online edition.}
	\label{fig:figure1}
\end{figure}

Figure A3 illustrates the $\dot M$ evolution of the most massive protobinary system and the most massive isolated protostar in models M4a and M8a in the top and bottom panels, respectively. In model M4a for the Kolmogorov-type turbulence (top panel), the primary component shown in blue forms at $t$ = 36.897 kyr while the secondary in red is created earlier at $t$ = 36.559 kyr. The first accretion burst for the primary companion occurs as $\dot M_{\rm prim}$ = 2.6 $\times 10^{-3}$ M$_{\odot}$ yr$^{-1}$ at $t$ = 36.897 kyr. The first accretion burst for the secondary companion occurs as $\dot M_{\rm sec}$ = 1.6 $\times 10^{-3}$ M$_{\odot}$ yr$^{-1}$ at $t$ = 36.643 kyr. The primary companion keeps evolving further with a general declining trend in $\dot M$ and randomly exhibits small accretion bursts until t $\approx$ 37.153 kyr. The secondary companion, after $t$ = 36.982 kyr, shows activity in $\dot M$ with its second and even more intense accretion burst of 2.0 $\times 10^{-3}$ M$_{\odot}$ yr$^{-1}$, which is then followed by a series of relatively stronger accretion bursts of short intervals of nearly 1 kyr. More interestingly, both the primary and secondary companions exhibit another more intense accretion bursts of $\dot M_{\rm prim}$ = 4.1 $\times 10^{-3}$ M$_{\odot}$ yr$^{-1}$ and $\dot M_{\rm sec}$ = 1.4 $\times 10^{-3}$ M$_{\odot}$ yr$^{-1}$ at t $\approx$ 37.153 kyr. The protobinary system beyond this point in time keeps showing regular events of accretion bursts such that both companions indicate mutually consistent intensity in $\dot M$ until we terminate the simulation at $t$ = 37.4 kyr. The isolated protostar in this model appears at $t$ = 36.643 kyr and shows a peak in $\dot M$ as $\dot M_{\rm iso}$ = 1.8 $\times 10^{-3}$ M$_{\odot}$ yr$^{-1}$ at $t$ = 36.728 kyr. After this, the $\dot M$ evolution shows a gradually declining trend and $\dot M$ drops down to $\dot M_{\rm iso}$ = 1.0 $\times 10^{-5}$ M$_{\odot}$ yr$^{-1}$ by the time we terminate the simulation.
In model M8a for the Burgers-type turbulence (bottom panel), the primary component shown in blue forms at $t$ = 43.351 kyr while the secondary in red is created at $t$ = 42.072 kyr. We suspect that the interaction between the two companions begins at t $\approx$ 46.773 kyr. From this time onward, the previously declining trend in the evolution of $\dot M$ for the two companions changes and starts to show some activity. For the primary companion, the most intense accretion burst of $\dot M_{\rm prim}$ = 3.4 $\times 10^{-3}$ M$_{\odot}$ yr$^{-1}$ occurs at $t$ = 46.881 kyr. At the same point in time, in comparison to the primary, the accretion rate $\dot{M}$ of the secondary shows an order of magnitude higher accretion burst of $\dot M_{\rm sec}$ = 1.0 $\times 10^{-4}$ M$_{\odot}$ yr$^{-1}$. This then followed by a series of successive mutually consistent accretion bursts which remain a constant feature during the later evolution of the two companions until we terminate the simulation at $t$ = 53.3 kyr. Interestingly, the secondary remains more active than the primary in its accretion bursts intensity. On average, these accretion bursts exhibit nearly an order of magnitude fluctuations in $\dot M$ with a frequency of nearly 1 kyr. The isolated protostar in this model appears at $t$ = 51.168 kyr. From its creation it shows a continuous decline and the mass accretion rate drops down to $\dot M_{\rm iso}$ = 1.0 $\times 10^{-5}$ M$_{\odot}$ yr$^{-1}$ at the time when we terminate the simulation.

Figures A4$-$A6 show for the case of the second random number seed, a more or less similar trends in the  evolution of $\dot M$ for the most massive protobinary system and the most massive isolated protostar in the respective models. The secondary companion remains more active than the primary and the spikes in the mass accretion rate remain evident during the later phase of the protobinary evolution. Also, the protobinary systems emerging from turbulence with a Kolmogorov-type turbulence exhibit more intense episodes of accretion than those which are formed with a Burgers-type turbulence.

\begin{figure}
	
	\centering
	\includegraphics[width=\columnwidth]{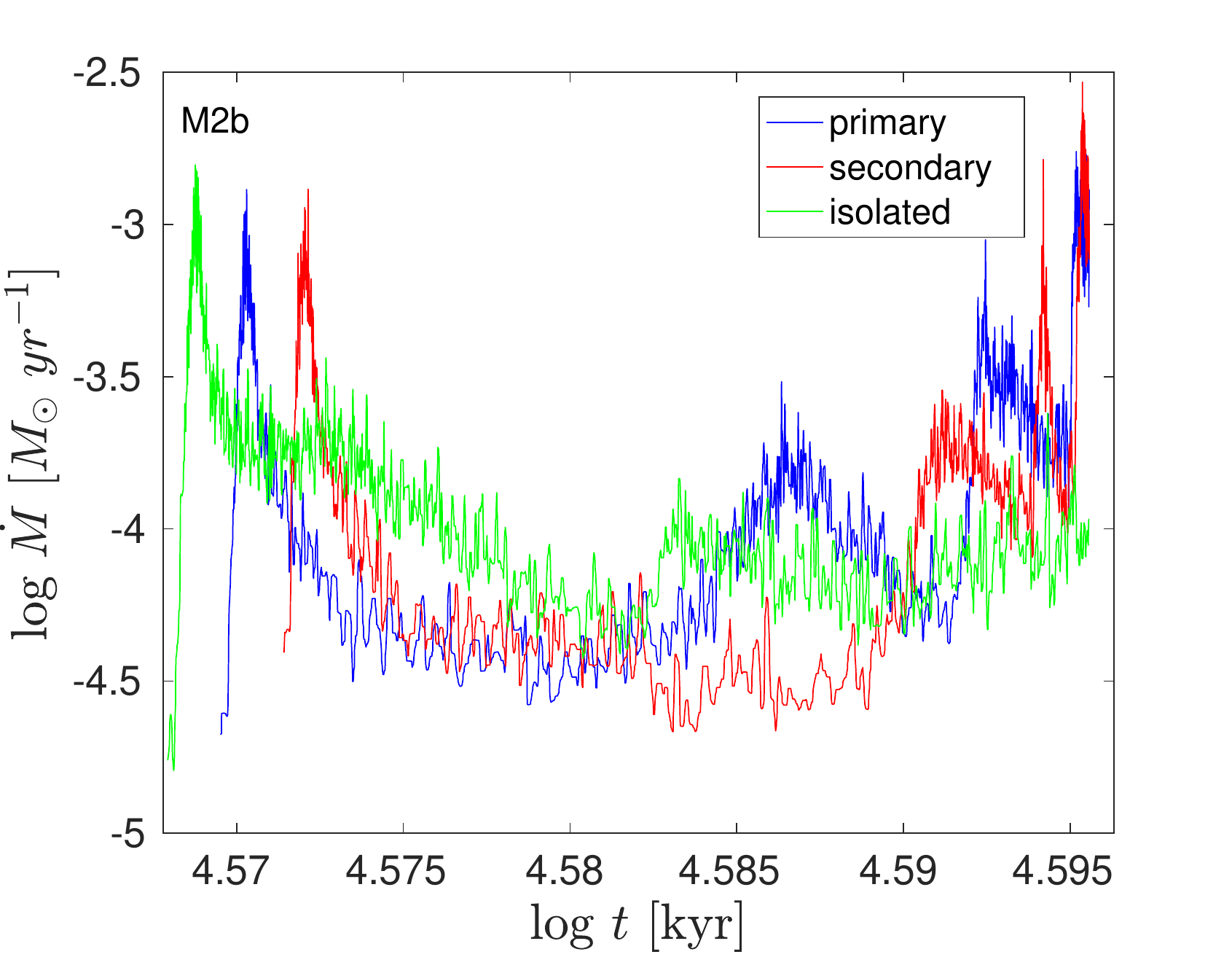}
	\includegraphics[width=\columnwidth]{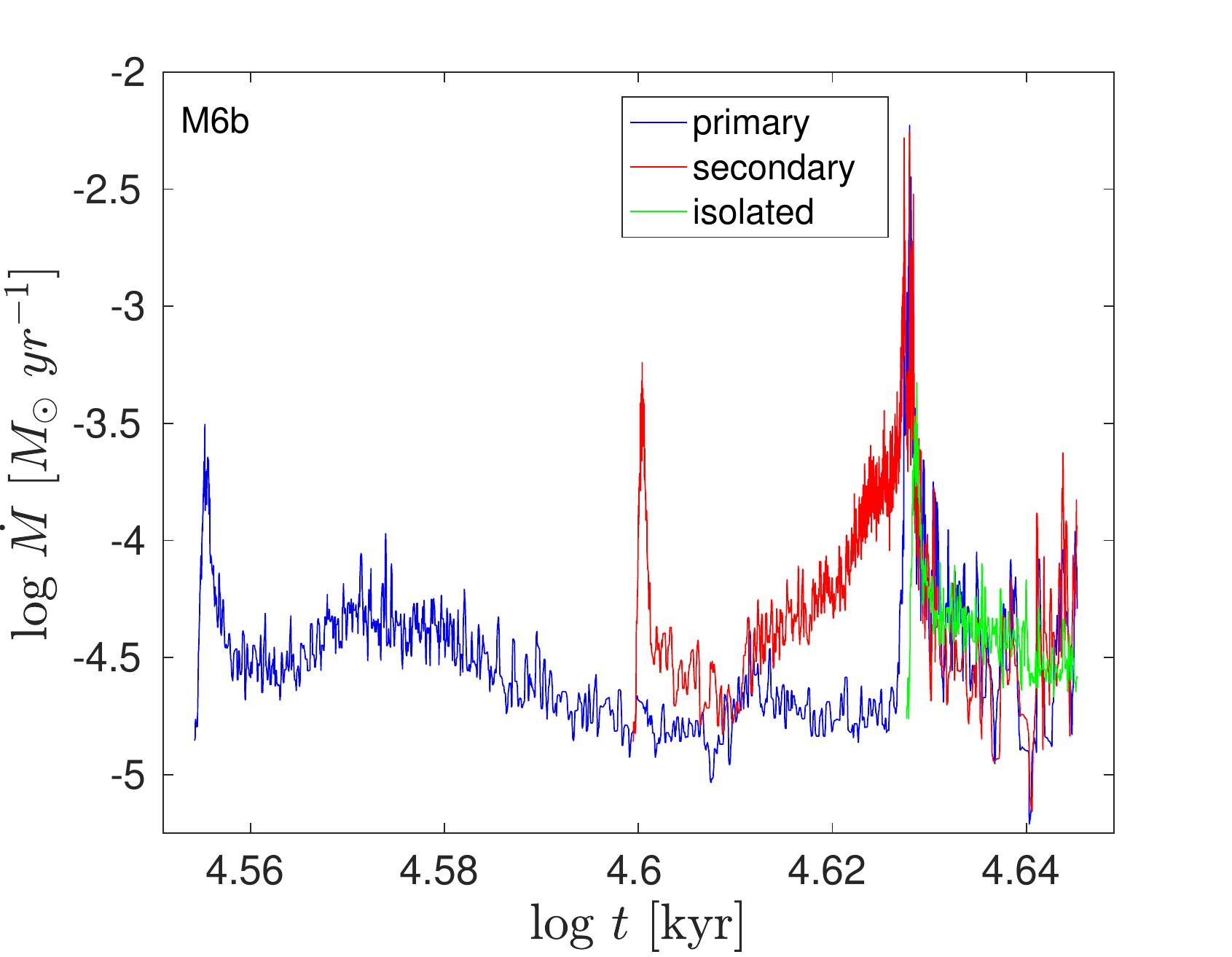}
	\caption{Accretion rates for the primary (blue), secondary (red), and isolated protostars (green) in models M2b (top panel) and M6b (bottom panel). The accretion rate is given in units of M$_{\odot}$ yr$^{-1}$ and the time is in kyr. Colour in online edition.}
	\label{fig:figure1}
\end{figure}

\begin{figure}
	
	\centering
	\includegraphics[width=\columnwidth]{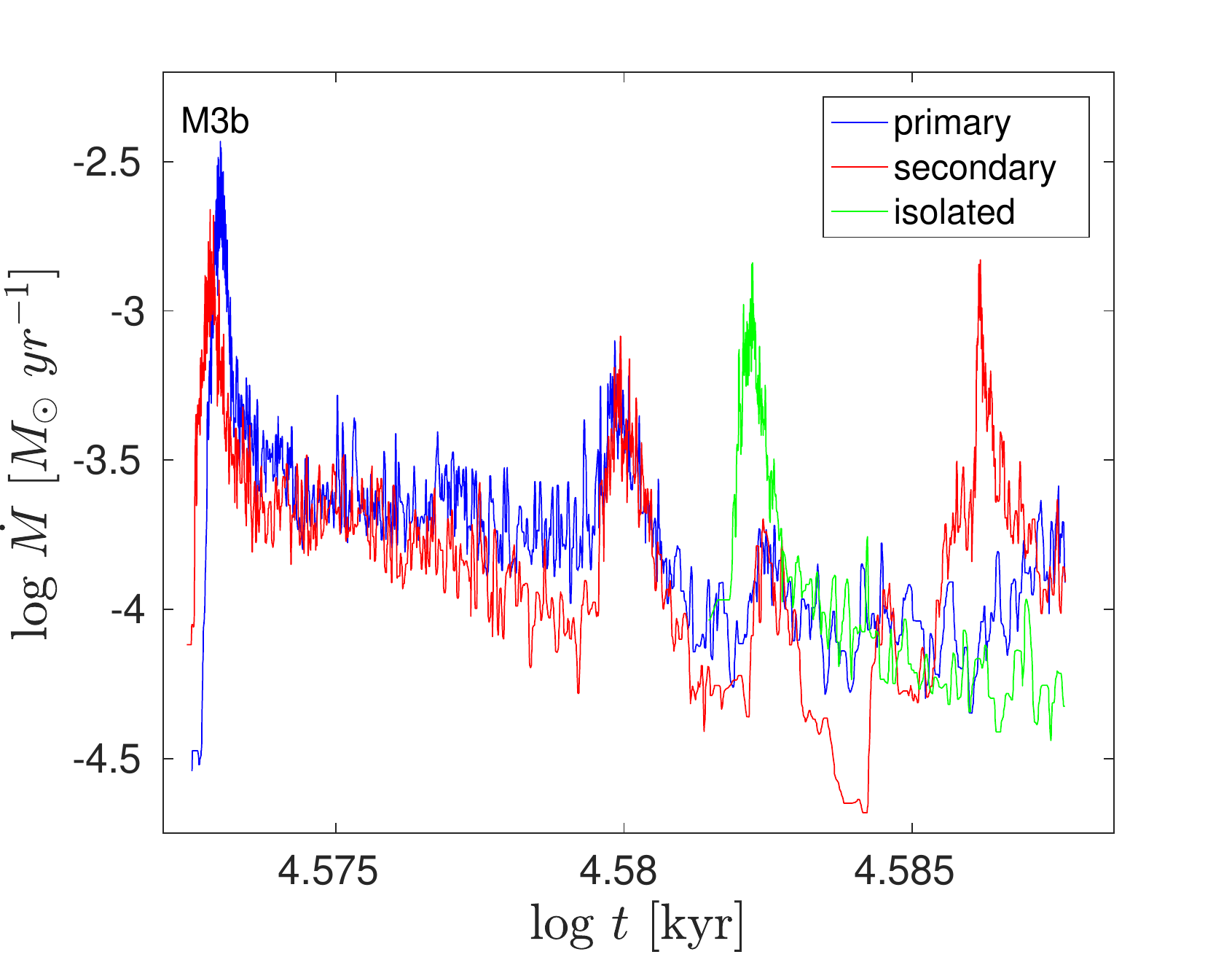}
	\includegraphics[width=\columnwidth]{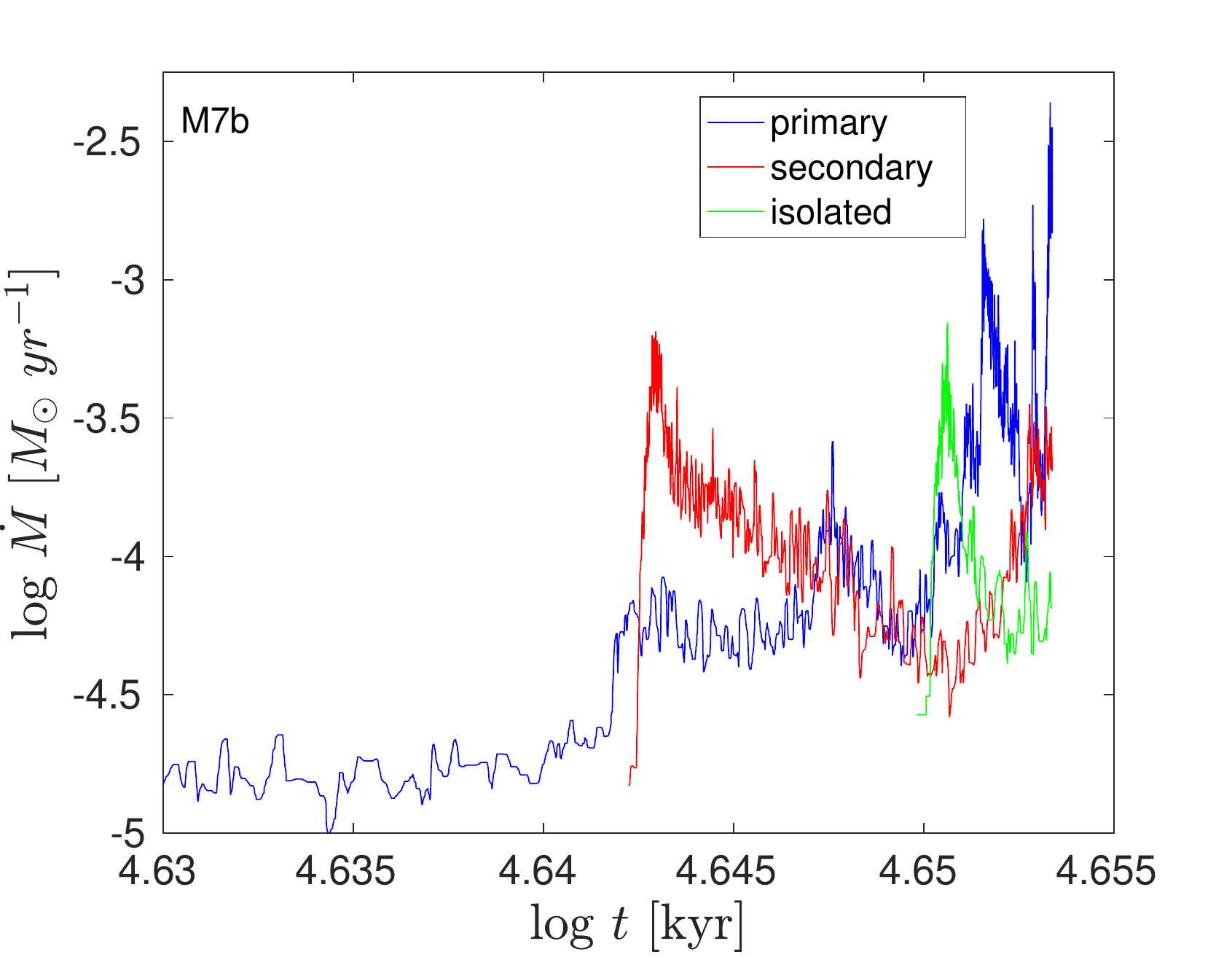}
	\caption{Accretion rates for the primary (blue), secondary (red), and isolated protostars (green) in models M3b (top panel) and M7b (bottom panel). The accretion rate is given in units of M$_{\odot}$ yr$^{-1}$ and the time is in kyr. Colour in online edition.}
	\label{fig:figure1}
\end{figure}
\begin{figure}
	
	\centering
	\includegraphics[width=\columnwidth]{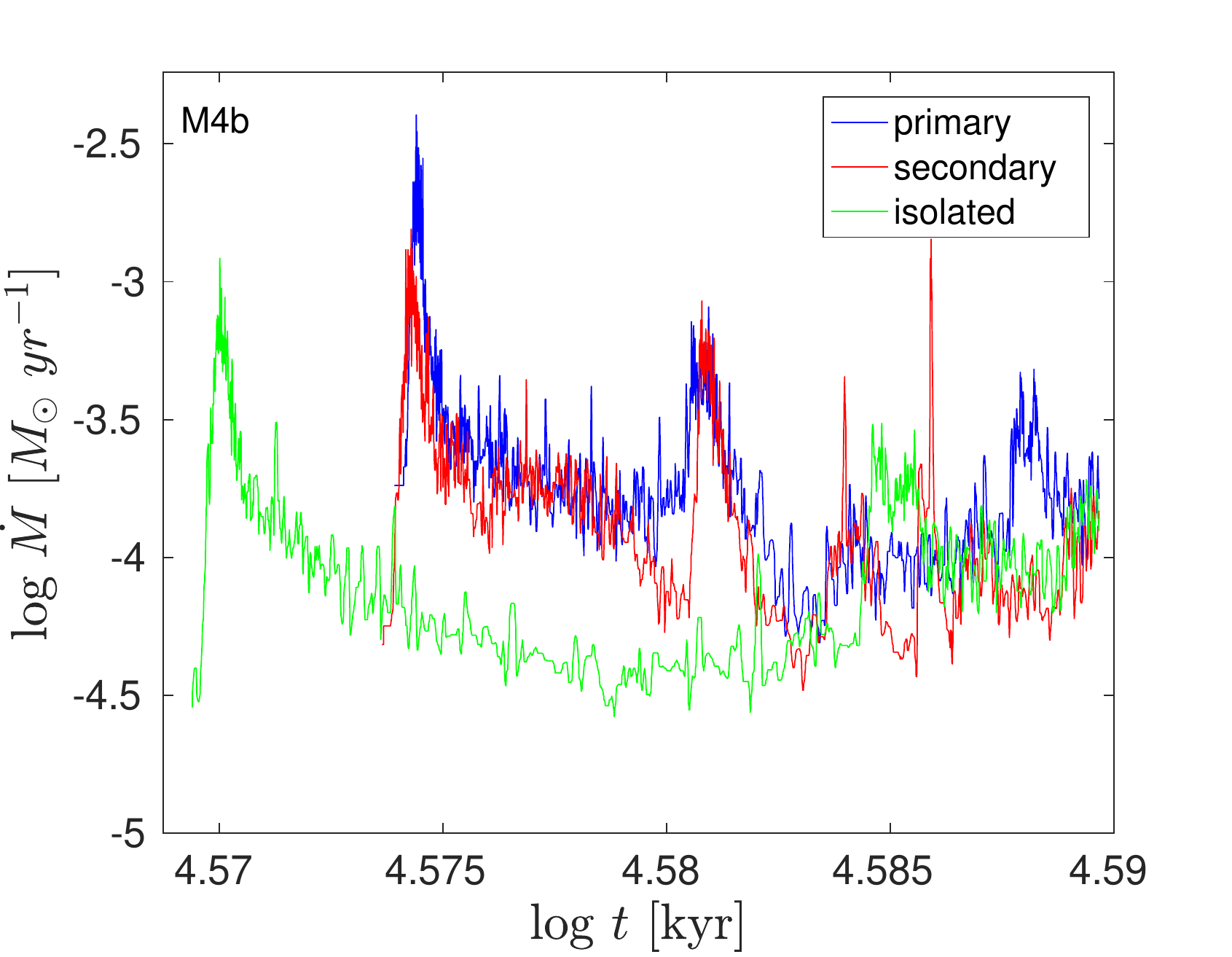}
	\includegraphics[width=\columnwidth]{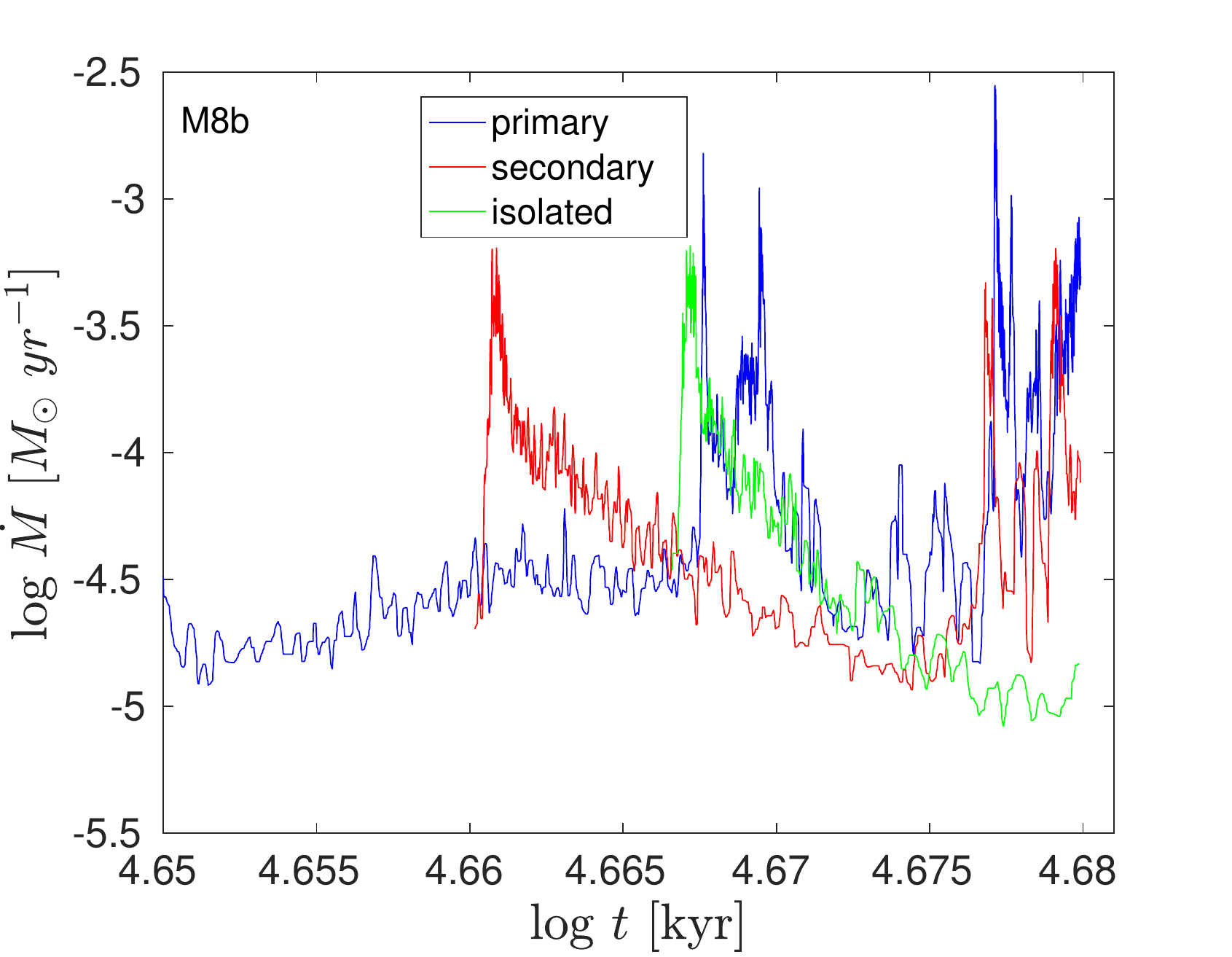}
	\caption{Accretion rates for the primary (blue), secondary (red), and isolated protostars (green) in models M4b (top panel) and M8b (bottom panel). The accretion rate is given in units of M$_{\odot}$ yr$^{-1}$ and the time is in kyr. Colour in online edition.}
	\label{fig:figure1}
\end{figure}


\bsp	
\label{lastpage}
\end{document}